\documentclass[11pt,a4paper]{article}
\pdfoutput=1

\usepackage{jcappub}
\usepackage{latexsym}
\usepackage{graphicx}
\usepackage{mathrsfs}
\usepackage{amsmath}
\usepackage{amsfonts} 
\usepackage{amssymb}  
\usepackage[bf, sf, FIGTOPCAP, nooneline, tight]{subfigure}
\usepackage{multirow}
\usepackage{multicol}
\usepackage[numbers,sort&compress]{natbib}
\usepackage{array}
\usepackage{bigstrut}
\usepackage{rotating}
\usepackage{booktabs}
\usepackage{hhline}

\def\urltilda{\kern -.15em\lower .7ex\hbox{\~{}}\kern .04em}

\def\tab#1{Table~\ref{#1}}

\def\fig#1{Fig.~\ref{#1}}
\def\figs#1#2{Figs.~\ref{#1} and \ref{#2}}

\def\sec#1{Sec.~\ref{#1}}
\def\eq#1{Eq.~\ref{#1}}

\definecolor{gold}{rgb}{0.85,.66,0}

\newcommand{\newc}{\newcommand}

\newc{\be}{\begin{equation}}
\newc{\ee}{\end{equation}}
\newc{\nn}{\nonumber}
\newc\ps{\mbox{ ps}}
\newc{\mev}{\mbox{ MeV}}
\newc{\gev}{\mbox{ GeV}}
\newc{\tev}{\mbox{ TeV}}
\newc{\GeV}{\gev}
\newc{\MeV}{\mev}
\newc{\TeV}{\tev}
\newc{\cl}{\text{CL}}
\newc\BR{BR}
\newc{\alphaemmz}{\alpha_{\text{em}}(m_Z)^{\overline{MS}}}
\newc{\alphas}{\alpha_s(m_Z)^{\overline{MS}}}
\newc\zetah{\zeta_h}
\newc\eg{{\rm {e.g.}}}
\newc\etal{{\rm {et al.}}}
\newc\ie{{\rm i.e.}}
\newc\etc{{\rm {etc}}}
\newc{\mhalf}{m_{1/2}}
\newc{\mzero}{m_0}
\newc{\tanb}{\tan\beta}
\newc{\azero}{A_0}
\newc{\sgn}{{\rm sgn}}
\newc{\deltaamususy}{\delta a_{\mu}^{\text{SUSY}}}
\newc{\bsg}{\bsgamma}
\newc\gmtwo{(g-2)_{\mu}}
\newc\deltaamu{\Delta a_{\mu}}
\newc{\abundchi}{\Omega_{\chi} h^2}
\newc{\msbar}{\overline{MS}}
\newc{\mtop}{m_t}
\newc{\mtpole}{m_t}
\newc{\hl}{h}
\newc{\mhl}{m_{\hl}}  
\newc{\mgut}{M_{\rm GUT}}
\newc{\mplanck}{M_{\rm P}}
\newc{\mpl}{M_{\text{Pl}}}
\newc{\msusy}{M_{\rm SUSY}}
\newc{\ms}{M_{\text{S}}}
\newc{\VEV}[1]{\langle #1 \rangle}
\newc{\sineff}{\sin^2 \theta_{\rm{eff}}}
\newc\MN{{\sf {MultiNest}}}

\newc\bsgamma{b\rightarrow s \gamma }
\newc\brbsgamma{\BR(\overline{B}\rightarrow X_s\gamma)}
\newc\bsmumu{\overline{B}_s\to\mu^+\mu^-}
\newc\brbsmumu{\BR(\overline{B}_s\to\mu^+\mu^-)}
\newc\bdmmumu{\overline{B}_d\to\mu^+\mu^-}
\newc\bbbarmix{\overline{B}_s\mbox{--}B_s}
\newc\delmbs{\Delta M_{B_s}}
\newc\brbtaunu{\BR(\overline{B}_u\to \tau \nu)}
\newc{\mbmbmsbar}{m_b(m_b)^{\msbar} }
\newc{\dRdE}{\frac{dR}{dE}}

\newc\AIPCP[3] {{AIP Conf. Proc.} {\bf #1} (#2) #3}
\newc\AJ[3] {{Astrophys. J.} {\bf #1} (#2) #3}
\newc\AJP[3] {{Am. J. Phys.} {\bf #1} (#2) #3}
\newc\AMS[3] {{Ann. Math. Statist.} {\bf #1} (#2) #3}                
\newc\AP[3] {{Ann. Phys.} {\bf #1} (#2) #3}
\newc\APJ[3] {{Astropart. J.} {\bf #1} (#2) #3}
\newc\APP[3] {{Astropart. Phys.} {\bf #1} (#2) #3}
\newc\APS[3] {{Astrophys. J. Suppl.} {\bf #1} (#2) #3}
\newc\ARNPS[3] {{Ann. Rev. Nucl. Part. Sci.} {\bf C#1} (#2) #3}
\newc\BA[3] {{Bayesian Anal.} {\bf C#1} (#2) #3}              
\newc\CPC[3] {{Comput. Phys. Commun.} {\bf C#1} (#2) #3}
\newc\CP[3] {{Contemp. Phys.} {\bf #1} (#2) #3}                     
\newc\EPJ[3] {{Euro. Phys. Journ.} {\bf C#1} (#2) #3}
\newc\JCAP[3] {{JCAP} {\bf #1} (#2) #3}
\newc\JHEP[3] {{JHEP} {\bf #1} (#2) #3}
\newc\JPG[3] {{J. Phys.} {\bf G #1} (#2) #3}
\newc\IJMP[3] {{Int. J. Mod. Phys.} {\bf A #1} (#2) #3}
\newc\MNRAS[3] {{Mon. Not. Roy. Astron. Soc.} {\bf #1} (#2) #3}
\newc\MPL[3] {{Mod. Phys. Lett.} {\bf A #1} (#2) #3}
\newc\NAR[3] {{New Astron. Rev.} {\bf #1} (#2) #3}                  
\newc\NCA[3] {{Nuovo Cimento} {\bf #1} (#2) #3}
\newc\NIM[3] {{Nucl. Instrum. Methods} {\bf #1} (#2) #3}
\newc\NIMA[3] {{Nucl. Instrum. Methods} {\bf A #1} (#2) #3}
\newc\NAT[3] {{Nature} {\bf #1} (#2) #3}
\newc\NPB[3] {{Nucl. Phys.} {\bf B #1} (#2) #3}
\newc\NPA[3] {{Nucl. Phys.} {\bf A #1} (#2) #3}
\newc\NPPS[3] {{Nucl. Phys. Proc. Suppl.} {\bf #1} (#2) #3}                
\newc\PLB[3] {{Phys. Lett.} {\bf B #1} (#2) #3}
\newc\PR[3] {{Phys. Rep.} {\bf #1} (#2) #3}
\newc\PRL[3] {{Phys. Rev. Lett.} {\bf #1} (#2) #3}
\newc\PRD[3] {{Phys. Rev.} {\bf D #1} (#2) #3}
\newc\PRC[3] {{Phys. Rev.} {\bf C #1} (#2) #3}
\newc\PTP[3] {{Prog. Theor. Phys.} {\bf #1} (#2) #3}
\newc\RMP[3] {{Rev. Mod. Phys.} {\bf #1} (#2) #3 }
\newc\RPP[3] {{Rept. Prog. Phys.} {\bf #1} (#2) #3 }
\newc\SC[3] {{Science} {\bf #1} (#2) #3 }
\newc\ZPC[3] {{Z. Phys.} {\bf C #1} (#2) #3}
\newc\Err[3] {{Erratum-ibid.} {\bf #1} (#2) #3 }

\title{Statistical coverage for supersymmetric parameter estimation: a case study with direct detection of dark matter}

\author[a]{Yashar Akrami,}
\author[a]{Christopher Savage,}
\author[a,b]{Pat Scott,}
\author[a]{Jan Conrad}
\author[a]{and Joakim Edsj\"o}

\affiliation[a]{Oskar Klein Centre for Cosmoparticle Physics and Department of Physics,\\
Stockholm University,\\
AlbaNova University Centre, SE-10691 Stockholm, Sweden}
\affiliation[b]{Department of Physics, McGill University,\\
3600 rue University, Montr\'eal, QC, H3A 2T8, Canada}

\emailAdd{yashar@fysik.su.se}
\emailAdd{savage@fysik.su.se}
\emailAdd{patscott@physics.mcgill.ca}
\emailAdd{conrad@fysik.su.se}
\emailAdd{edsjo@fysik.su.se}

\abstract{Models of weak-scale supersymmetry offer viable dark matter (DM) candidates. Their parameter spaces are however rather large and complex, such that pinning down the actual parameter values from experimental data can depend strongly on the employed statistical framework and scanning algorithm. In frequentist parameter estimation, a central requirement for properly constructed confidence intervals is that they cover true parameter values, preferably at exactly the stated confidence level when experiments are repeated infinitely many times. Since most widely-used scanning techniques are optimised for Bayesian statistics, one needs to assess their abilities in providing correct confidence intervals in terms of the statistical coverage. Here we investigate this for the Constrained Minimal Supersymmetric Standard Model (CMSSM) when only constrained by data from direct searches for dark matter. We construct confidence intervals from one-dimensional profile likelihoods and study the coverage by generating several pseudo-experiments for a few benchmark sets of pseudo-true parameters. We use nested sampling to scan the parameter space and evaluate the coverage for the benchmarks when either flat or logarithmic priors are imposed on gaugino and scalar mass parameters. The sampling algorithm has been used in the configuration usually adopted for exploration of the
Bayesian posterior. We observe both under- and over-coverage, which in some cases vary quite dramatically when benchmarks or priors are modified. We show how most of the variation can be explained as the impact of explicit priors as well as sampling effects, where the latter are indirectly imposed by physicality conditions. For comparison, we also evaluate the coverage for Bayesian credible intervals, and observe significant under-coverage in those cases.}


\begin{document}

\maketitle

\section{Introduction} \label{sec:intro}

Over the past few years, our knowledge about the Universe and its basic constituents has been significantly enhanced. Thanks to various high-precision astrophysical and cosmological observations, we have been able to construct a standard and very successful theoretical model of our Universe (for an introduction, see Refs.~\cite{cosmology:Weinberg,cosmology:Mukhanov}).

This coherent picture, however, requires introducing new ingredients to our understanding of the Universe's elementary building-blocks and their interactions. If our current standard model of cosmology is a correct description of Nature, we need, in particular, to identify the nature of its two essential elements, namely dark matter (DM) and dark energy (DE), which together make up more than $95\%$ of the entire energy budget of the Universe (see e.g. Refs.~\cite{Komatsu:2008hk,Iocco:2008va,Riess:2004nr,Astier:2005qq,Cole:2005sx,Kessler:2009ys,Massey:2007wb}).

As far as the DM problem is concerned, there has recently been remarkable progress towards the understanding of its nature (for some reviews, see e.g. Refs.~\cite{DMBergstrom:2000,DMBertone:2005,DMBergstrom:2009,DMBertone:2010}). From the theoretical point of view, there is indeed no lack for number or diversity in viable DM candidates. Perhaps the dominant paradigm is the conjecture that DM is composed of weakly interacting massive particles (WIMPs). Many theories beyond the standard model (SM) of particle physics offer such particles as their natural ingredients. The lightest neutralino or gravitino in supersymmetric (SUSY), and the lightest Kaluza-Klein (KK) particle in extra-dimensional extensions are some examples. These models however still require further experimental constraint, or verification.

In addition to cosmological observations, which deal with pure gravitational properties of DM particles, there are several other experiments that can probe additional characteristics of DM. These are mainly divided into three categories: direct detection (DD; where a WIMP is sought by its interaction with atomic nuclei of some target material), indirect detection (ID; where signals from WIMP self-annihilation in astrophysical sources are sought in gamma- or other cosmic rays), and accelerator searches (where WIMPs are expected to be produced through high energy processes that happen at particle colliders). Needless to say, the nature of DM can be claimed as having been eventually identified within a specific theoretical framework only if the results of all different experiments agree with a high level of confidence.

Perhaps the most popular model for DM is that of SUSY, in particular, in the framework of the Minimal Supersymmetric Standard Model (MSSM) when the lightest SUSY particle (LSP) is the lightest neutralino (for a general introduction to softly-broken weak-scale supersymmetry, see Refs.~\cite{Martin:9709356,SUSY:Aitchison,SUSY:Baer}, and for an introduction to supersymmetric DM in particular, see Ref.~\cite{DMJungman}). The MSSM assumes conservation of $R$-parity, making the LSP absolutely stable. If the neutralino (which is electrically neutral and weakly interacting) is the LSP, it can be considered as an interesting candidate for cold DM.

The MSSM however contains more than a hundred free parameters which are highly unconstrained. There have been several attempts at comparing MSSM predictions with observation to pin down the favoured values of the parameters, but every such inference has turned out to depend strongly on the scanning technique, statistical framework and measure, and our prior knowledge about the employed parameterisation and the actual values of the parameters (see e.g. Refs.~\cite{Allanach:07050487,Trotta:08093792,Akrami:2009hp,Scott:2009jn} and references therein). Even after a dramatic reduction of the number of model parameters, e.g. in models like minimal supergravity (mSUGRA)~\cite{mSUGRA} or the constrained MSSM (CMSSM)~\cite{CMSSM}, the parameter space is still very complex such that many phenomenological analyses of the model are still far from being conclusive. The situation will improve though when more data become available in the future, for example, by the Large Hadron Collider (LHC)~\cite{Roszkowski:2009ye,Bertone:2010rv} or more powerful DD experiments~\cite{Akrami:2010,Bertone:2010rv}.  

In this paper, we highlight yet another problem in SUSY parameter estimation and statistical inference: the degree to which stated confidence/credible intervals on fitted model parameters do in fact contain the intended amount of probability. This is a statistical issue that needs to be adequately controlled in order to make robust statements about any theoretical model, in particular, SUSY models in the context of DM searches. We demonstrate the problem through a case study where some data from a typical DD experiment are employed and the analysis is performed using a state-of-the-art and widely-used scanning technique optimised for Bayesian inference. 

In~\sec{sec:problem}, we introduce the problem with a brief review of Bayesian and classical (frequentist) statistics, and after describing our specific example in~\sec{sec:case}, we present and discuss our results in~\sec{sec:results}. We summarise and conclude in~\sec{sec:concl}.

\section{The problem} \label{sec:problem}
\subsection{Bayesian and frequentist approaches to SUSY parameter estimation} \label{sec:approaches}

Suppose that we have a theoretical model with $m$ free parameters $\Theta=(\theta_1,\theta_2,...,\theta_m)$ and also suppose that the model has different physical predictions corresponding to each particular set of parameters $\Theta$. In general, these predictions can be of two types: (1) they are physical observables with fixed values that can be computed theoretically, or (2) they are observables that are stochastic by nature (because the physical processes involved in their predictions are e.g. thermodynamic or quantum mechanical). In the latter case, the intrinsic stochasticity of the observable implies that the theoretical predictions of the model at hand are given in terms of probability distribution functions rather than fixed quantities. However, even in cases where the predicted quantity has a fixed value, there are various extrinsic sources of uncertainty that make a statistical analysis of the model unavoidable. These uncertainties can be in both predicted and measured values of the observables. Theoretical uncertainties come from various approximations one makes in deriving the theoretical values from the underlying model, and experimental uncertainties are caused by statistical and systematic errors associated with imperfection of the corresponding experimental apparatus. Another source of uncertainty is our ignorance of potential backgrounds and this can be estimated either theoretically or experimentally. Let us now assume that $n$ such quantities exist with experimentally measured values of $D=(d_1,d_2,...,d_n)$. The question then becomes: how can one extract useful information about the actual values of the model parameters $\Theta$ by looking at the measured observables $D$, taking into account all sources of errors and uncertainties?

Statisticians' answers to this question fall primarily into two categories, which have to do with two conceptually very different definitions of probability and how it is assigned to parameters and measurements: Bayesian and frequentist (classical) statistics (for a detailed discussion, see e.g. Ref.~\cite{Cowan:1998}).

Suppose the probability of obtaining the data set $D$ for a particular set of theoretical parameters $\Theta$ by making some measurements is given in terms of the probability density function (PDF) $P(D|\Theta)$ (where $\int P(D|\Theta) dD = 1$). This is assumed to include all possible contributions to our uncertainty about the measurements. From the point of view of a Bayesian statistician, it is perfectly rational to also speak of $P(\Theta|D)$, i.e. the PDF associated with the probability of the particular set of model parameters $\Theta$ being `true' if the particular set of experimental data $D$ is obtained (for an introduction to general applications of Bayesian statistics
in physics, see e.g. Ref.~\cite{DAgostini:1995}, and for reviews of its applications in cosmology, see e.g. Refs.~\cite{Trotta:2005,Trotta:2008,Liddle:2009,Hobson:2010}). This is because probability for a Bayesian indicates the degree of belief about a statement or hypothesis. The hypothesis in the case of parameter estimation is that a specific set of model parameters is the true set of values. In this case $\int P(\Theta|D) d\Theta = 1$. The two quantities $P(D|\Theta)$ and $P(\Theta|D)$ are then related by applying Bayes' Theorem
\be
\label{eq:bayestheorem}
P(\Theta|D)=\frac{P(D|\Theta)P(\Theta)}{P(D)}.
\ee
Here, $P(\Theta)$ represents our degree of belief about $\Theta$ being the true values of parameters, `prior to' our inclusion of the information given by the data $D$; $P(\Theta)$ is simply called the `prior'. $P(D)$ denotes the total probability of obtaining the particular set of data $D$ given the theoretical model and irrespective of the actual values of the model's parameters; $P(D)$ is called the `evidence'. Our updated knowledge about the values of model parameters is all embodied in the quantity $P(\Theta|D)$ on the left hand side of the equation, which is called the `posterior' PDF. The ultimate objective in any Bayesian analysis of a model is then to accurately compute and map the posterior as a function of $\Theta$. In this case, $P(D|\Theta)$ on the right hand side (which is a PDF in the data space) is considered as a function of $\Theta$ and is known as the `likelihood function'; we denote it accordingly $\mathcal{L}(\Theta)$.

One major issue for Bayesian parameter estimation is how to choose the prior $P(\Theta)$. Although it reflects our prejudice about the parameter values, one may think that the `natural' choice for being unbiased in this respect is to work with uniform (or flat) priors for all parameters. This however does not help, because it depends upon the way the model is parameterised, or equivalently upon the metric defined on the parameter space. An alternative way to choose the priors, which is perfectly acceptable and actually strongly recommended, is to use the posterior from a previous inference cycle as the prior for the next. One should however notice that the issue still exists for the initial choice of the priors in this inference chain. In cases where a large number of inference cycles are used or the data are constraining enough, the impact of the initial priors can become weak or negligible. In existing statistical analyses of SUSY models, the effects of priors are very strong compared to data, making their selection an important issue.

To understand when the effects of priors become important, we should compare the value of the likelihood $\mathcal{L}(\Theta)$ with that of the prior at all $\Theta$; the data are labelled `constraining' if the likelihood dominates and `unconstraining' if priors do.

Contrary to Bayesian statistics, frequentism defines the probabilities in a manifestly `objective' way. Suppose that we have a repeatable experiment which gives a set of experimental outcomes. The probability of a certain outcome $D$ is then defined as the fraction of times that $D$ happens when the experiment is repeated an infinite number of times such that every time the repetition is equiprobable. Given a point $\Theta$ in the model parameter space, this probability is equal to the likelihood $\mathcal{L}(\Theta)$. Since one cannot assign objective probabilities to the model parameters in a similar way, the posterior PDF $P(\Theta|D)$ is meaningless and the only well-defined quantity in this framework is the likelihood $\mathcal{L}(\Theta)$.

\subsection{Credible intervals, confidence intervals and statistical coverage} \label{sec:coverage}
The central task of any statistical parameter estimation procedure is, in addition to giving an estimate of the value for a model parameter, to tell us how well the parameter is estimated, i.e.~to provide uncertainties in the determined value. This is performed differently in Bayesian and frequentist frameworks by providing the so-called credible and confidence regions (or intervals), respectively.

In Bayesian statistics, a direct corollary of~\eq{eq:bayestheorem} is that if we are interested in joint posterior PDFs for some parameters or the one-dimensional (1D) PDF for a single parameter, they can be obtained by simply integrating (or `marginalising') over the other parameters. For example, the 1D `marginal' posterior PDF for the parameter $\theta_i$ is
\be \label{eq:margpdf} \mathbb{P}(\theta_i|D)=\int
P(\Theta|D)d\theta_1...d\theta_{i-1}d\theta_{i+1}...d\theta_m.
\ee
Suppose that we are now interested in 1D intervals for a parameter $\theta_i$ corresponding to a confidence level $\alpha$.\footnote{Strictly speaking, this should be called a `credibility level' for Bayesian statistics, but we adhere to `confidence level' here since the term is widely used in the literature also for Bayesian inference~\cite{Cowan:1998,Feldman:1998,pdg07}.} This is performed in Bayesian inference by finding an interval $[\theta_i^1,\theta_i^2]$ for which
\be \label{eq:credinterval} \int_{\theta_i^1}^{\theta_i^2}\mathbb{P}(\theta_i|D)d\theta_i=\alpha.
\ee
These intervals are called `credible intervals'. Clearly, there are an infinite number of intervals that can be constructed given this prescription. One can specify the interval by for example constructing the shortest interval, the symmetric interval around the posterior mean, or just the upper/lower limits on the parameter value.

In frequentist inference, on the other hand, where no marginal posterior $\mathbb{P}(\theta_i|D)$ exists, one follows the Neyman definition of `confidence intervals' which are constructed only from the likelihood function $\mathcal{L}(\Theta)$~\cite{Neyman}. The procedure is as follows: suppose that a confidence interval $[\theta_i^1,\theta_i^2]$ is constructed such that it corresponds to a confidence level $\alpha$; $\theta_i^1$ and $\theta_i^2$ are functions of the measured experimental data $D$ and the true parameter is believed to reside in the interval. The requirement for this interval to be constructed properly is that if the experiments are repeated $N$ times and new intervals are reconstructed using the same procedure as the original one, these intervals contain the true value $M$ times such that
\be \label{eq:confinterval}  \lim_{N \to \infty}\frac{M}{N}=\alpha. 
\ee
One immediate consequence of the Neyman procedure for constructing frequentist confidence intervals is the requirement that~\eq{eq:confinterval} should hold for any hypothetical true values of model parameters. That is, if we assume that any arbitrary values are chosen by Nature, simulate several pseudo-experiments based on the PDF $P(D|\Theta)$, and work out the confidence intervals corresponding to a confidence level $\alpha$ for the entire ensemble of experiments, we expect to `cover' the pseudo-true values in a fraction $\alpha$ of cases. This requirement for correct 
`coverage' serves as a powerful measure of success for any statistical method in extracting frequentist confidence intervals (for a discussion on statistical coverage, see e.g. Ref.~\cite{Feldman:1998}).

Obviously, a perfect construction of confidence intervals gives exact coverage. Approximate methods however can give rise to both `under'- and `over'-coverage. Under-coverage at a given confidence level $\alpha$ means that at least for one parameter, the fraction of realisations in which the pseudo-true value is contained in the corresponding confidence interval is smaller than $\alpha$. Over-coverage is defined in an analogous manner but for fractions larger than $\alpha$.

Although both under- and over-coverage are problematic, the former is much more severe; it shows that our estimates of the true values for model parameters cannot be trusted since the true parameters may be located anywhere outside the intervals. Over-covering intervals, on the other hand, simply provide conservative estimates of the parameters.

One frequentist technique which provides correct coverage at a desired confidence level is the so-called `confidence belt' construction proposed by Feldman \& Cousins~\cite{Feldman:1998}. This technique however requires the full power of Neyman construction which is rather hard to implement numerically, especially for models with a large number of parameters.

Therefore in practice, instead of a full Neyman construction of the confidence intervals, one employs some approximate methods to estimate the intervals in a numerically feasible and sufficiently accurate way. Most of these methods are based on assuming a Gaussian shape for the likelihood function in both data and parameter spaces, an assumption that breaks down for cases with complex parameter spaces and insufficient data (i.e. low statistics). It is important to note that for cases with high statistics, the likelihood becomes Gaussian in the data, where it is a PDF, and not in the parameters where it is not a PDF. Therefore the approximate methods for constructing the confidence intervals do not necessarily become exact even in the limit of high statistics.

One of the various proposals for making the approximation, which is widely used in SUSY parameter extraction, is based on the `profile likelihood'. In this method, in order to make sensible inference about some parameters, one maximises (or `profiles') the likelihood along the unwanted parameters in the parameter space~\cite[and references therein]{profilelike}. As an example, the 1D profile likelihood $\mathbb{L}(\theta_i)$ for a single parameter $\theta_i$ becomes
\be
\label{eq:proflike}
\mathbb{L}(\theta_i)\equiv\max_{\theta_1,...,\theta_{i-1},\theta_{i+1},...,\theta_m}\mathcal{L}(\Theta).
\ee
To construct the confidence intervals, one first finds the point with the highest likelihood, and then gives uncertainties upon the parameter values by constructing iso-likelihood contours in the parameter space depending on the particular confidence level and the number of remaining free parameters after profiling over the unwanted ones.

It is important to remark that the inferences based on profile likelihoods and marginal posteriors (\eq{eq:margpdf}) do not agree in general, especially when we lack enough experimental data and the parameter space is complex. This is the case for SUSY models with currently available data.

In addition, as stated earlier in this section, if the likelihood has a Gaussian form in both data and parameters, all approximate methods for constructing confidence intervals, including the profile likelihood scheme, are expected to give exact coverage. This means that in general, one does not expect proper coverage for a complex parameter space even when sufficient data are available (i.e. for the case of high statistics). This is because although the likelihood becomes Gaussian in the data space, it might still be very different from a Gaussian function in the parameter space. Therefore, in general, for complex parameter spaces either with high or low statistics one usually expects poor coverage from the profile likelihood; it gives both under- and over-coverage~\cite{Feldman:1998}.

It should be noted that even in cases where the profile likelihood is expected to give correct coverage, it does so only if it is mapped accurately enough. However, most of the popular scanning algorithms employed for this purpose are designed and optimised for Bayesian analyses. These include Markov-Chain Monte Carlos (MCMCs) and methods based on nested sampling~\cite{SkillingNS1,SkillingNS2} such as~\MN~\cite{MultiNest1,MultiNest2}. Samples based on these techniques are affected by the choice of priors, which in turn impact the mappings for profile likelihoods. It is therefore interesting to know how good these techniques are in a frequentist framework, and one way of testing this is to study the degree of coverage one can achieve by employing them; this is the main goal of this paper.

On the other hand, for Bayesian credible intervals, one does not expect correct coverage since the definition of the intervals is established upon entirely different bases. However, it is probably true that most physicists would calibrate even Bayesian credible intervals using a frequentist approach, i.e. repeated experiments~\cite{Cousins:1995}. Therefore, it is interesting to also look at the credible intervals in terms of the statistical coverage.

\section{The case study: CMSSM scans and direct detection of dark matter} \label{sec:case}
\subsection{Theoretical setup and scanning strategy} \label{sec:setup}

The SUSY model we analyse here is the CMSSM~\cite{CMSSM}. Here, inspired by the simplest gravity-mediated SUSY-breaking mechanism called minimal supergravity (mSUGRA), and the natural connection between SUSY and grand unified theories (GUTs), all scalar, gaugino and trilinear mass parameters of the MSSM Lagrangian are unified at the gauge coupling unification scale ($\sim$10$^{16}$\,GeV). These are denoted by $\mzero$, $\mhalf$ and $\azero$, respectively. By requiring electroweak symmetry breaking (EWSB) to be fulfilled at low energies, the CMSSM possesses only five new free parameters compared to the SM,
\be \label{CMSSMparams}  \mzero,\mhalf,\azero,\tanb,\sgn\,\mu,
\ee
where $\tanb$ is the ratio of the up-type and down-type Higgs vacuum expectation values and $\mu$ is the Higgs/higgsino mass parameter in the MSSM superpotential. The magnitude of $\mu$ is fixed by the EWSB requirement and only its sign remains to be determined by experiment; this implies four continuous and one discrete parameters for the model.

In this paper, we fix $\mu$ to be positive, reducing the number of parameters by one. We instead add four SM quantities to our set of free parameters as `nuisance' parameters: the pole top quark mass ($\mtpole$), the bottom quark mass evaluated at its equivalent energy scale ($\mbmbmsbar$), and the electromagnetic and strong coupling constants both evaluated at the $Z$-boson pole mass ($\alphaemmz$ and $\alphas$, respectively). $\msbar$ stands for the modified minimal subtraction renormalisation scheme, and its appearance as superscripts means that the corresponding quantities are computed in that scheme. These nuisance parameters are the SM parameters with the largest uncertainties and strongest impacts upon CMSSM predictions.

The resulting eight-dimensional parameter space is used in this paper to study the statistical coverage. The general MSSM contains a much larger and more complex parameter space, but its analysis requires a huge amount of computational power. We have chosen the CMSSM as a testbed model to show the problem with coverage. We believe that the problem becomes more severe when the full MSSM is analysed.

One important ingredient of any coverage study is scanning over model parameters to construct confidence intervals.  This process should be performed several times. Simple scanning techniques, such as fixed grid or purely random scans turn out to be useless even for low-dimensional SUSY parameter spaces, including the CMSSM (For a recent review of different scanning techniques, see e.g. Ref.~\cite{Akrami:2009hp} and references therein). This means that one needs to work with more sophisticated scanning algorithms that are fast and efficient.

Most existing scans are based on either MCMCs or nested sampling.  These are both optimised for Bayesian scans (for an example of frequentist scanning techniques, see e.g. Ref.~\cite{Akrami:2009hp}). The two methods are shown to give very similar results, but the latter takes significantly less computational effort (see e.g. Ref.~\cite{Trotta:08093792}). We therefore work with nested sampling throughout this paper. The particular algorithm is called~\MN~\cite{MultiNest1,MultiNest2} and is implemented in \textsf{SuperBayeS}~\cite{Trotta:08093792,deAustri:2006pe,Roszkowski:2007fd} (available from Ref.~\cite{superbayes}), a publicly available package for exploring the CMSSM parameter space and comparing its predictions with different experimental data. \textsf{SuperBayeS} includes many standalone packages as submodules which calculate different quantities for a given point in the CMSSM parameter space. Packages that we have used in our analysis are \textsf{SOFTSUSY}~\cite{softsusy} (available from Ref.~\cite{softsusyweb}) for solving renormalisation group equations (RGEs) from the GUT scale down to the electroweak scale where observable quantities such as the neutralino mass are defined, and \textsf{DarkSUSY}~\cite{darksusy} (available from Ref.~\cite{darksusyweb}) for calculating various DM DD quantities that we use in constructing our likelihood function; we discuss this in the next section.
 
\subsection{Direct detection likelihood, benchmarks and pseudo-experiments} \label{sec:DD}

A central quantity in any statistical parameter scan is the likelihood function. \textsf{SuperBayeS} provides the possibility of combining likelihood contributions from experimental constraints on different types of observables. These include various existing collider and cosmological data with a potential for including DM direct and indirect detection constraints in a global-fit framework. The ideal strategy is then to incorporate all these data in our analyses. However, in order to evaluate the coverage, we need to simulate the considered experiments several times for each selected set of pseudo-true parameter values and then scan over the parameter space based on each set of generated data. This is a lengthy process and particularly cumbersome when contributions from certain observables (such as the DM relic density) are taken into account. Again, analogous to our strategy in selecting the model, we construct our likelihood function such that it reflects interesting properties of the model in terms of our particular problem, while not being overly computationally demanding. The choice we make is to work only with DM DD pseudo-data.

DD experiments aim to observe low energy nuclear recoils
induced by the scattering of relic neutralinos off nuclei in the
detector \cite{Goodman:1984dc,Smith:1988kw,Lewin:1995rx,DMJungman}.  For an
experiment that observes $N$ events with energies $E_i$, we define
\begin{equation} \label{eqn:DDLike}
   \mathcal{L}_{DD}(\Theta) \ = \ P(N|\mu(\Theta)) \prod_{i=1..N} \! f(E_i|\Theta)
\end{equation}
as the likelihood of the experimental result, where $\Theta$ denotes the CMSSM and SM nuisance parameters. Here, $P(N|\mu)$ is the
probability of seeing $N$ events for a Poisson distribution with
average $\mu$, while
\begin{equation} \label{eqn:fE}
  f(E) = \frac{1}{N_f} \dRdE \, , \qquad \textrm{with} \quad
  N_f \equiv \int_{E_1}^{E_2} dE \, \dRdE \, ,
\end{equation}
is the probability of an event having an observed energy $E$ over an energy range $E_1$--$E_2$, and $N_f$ is a normalisation factor. The quantity $\dRdE$ is the expected event spectrum after accounting
for various analysis cuts and incorporating the finite energy
resolution of the detector; see Ref.~\cite{Akrami:2010} for a description of
how it is calculated.  In principle, background processes can
contaminate the signal and contribute to the event spectrum $\dRdE$;
however, we do not include any background contributions here.
The expected number of events and energy spectrum depend on the
neutralino mass $m_{\tilde\chi^0_1}$, the spin-independent
neutralino-proton scattering cross-section $\sigma^{SI}_p$, and
the two spin-dependent neutralino-nucleon scattering cross-sections
$\sigma^{SD}_p$ \& $\sigma^{SD}_n$, each of which are fixed by the
parameters in a given CMSSM model.\footnote{
  A fourth cross-section $\sigma^{SI}_n$, for spin-independent
  neutralino-neutron scattering, is nearly identical to $\sigma^{SI}_p$
  in the CMSSM and is not treated as an independent parameter.}
For the DD experiment and SUSY models we consider in this
paper, spin-dependent interactions are not significant and will not
be further discussed (though they are included in the calculations).

We take for our hypothetical DD experiment a XENON10-like 
detector \cite{Aprile:2010bt} with an exposure of 1000~kg-days.
We assume an allowed event energy range of 2-75~keV, take efficiencies
from Ref.~\cite{Angle:2009xb}, and use an energy resolution of~\cite{Baudis:2008pc}
\begin{equation} \label{eqn:XENON10ER}
  \sigma(E) = (0.579\,\mathrm{keV}) \sqrt{E/\mathrm{keV}} + 0.021 E \, .
\end{equation}
The 2~keV threshold of the XENON10 analysis presented in Ref.~\cite{Angle:2009xb} was based upon the understanding of the XENON10 detector behaviour at the time.  However, more recent detector energy calibration measurements suggest
the actual XENON10 energy threshold is higher than 2~keV (see \eg\ 
Ref.~\cite{Savage:2010tg} and references therein) and the energy resolution
at low recoil energies is much more complicated than as described
above \cite{Sorensen:2010hq}.  However, we are simply using these
characteristics for a hypothetical experiment and these issues are
not relevant for the purposes of this paper.

With this choice of the likelihood function, we then need to choose some benchmark points in the parameter space representing the pseudo-true parameters for our coverage study. These points can in principle be selected arbitrarily since the coverage requirement should hold for all points regardless of being well-fitted or not. The ideal situation is to evaluate the degree of coverage corresponding to every single point in the parameter space, but due to our limitation in the computational power, we just pick a couple of points for our analysis. We call these points benchmark 1 and benchmark 2.~\tab{tab:params4BMs} shows values of all CMSSM and SM nuisance parameters at the two benchmarks, as well as the corresponding DD quantities $m_{\tilde\chi^0_1}$ and $\sigma^{SI}_p$.\footnote{We should remark that indeed our two benchmark points `1' and `2' have not been chosen completely arbitrarily. They correspond to the posterior mean and the best-fit (highest-likelihood) point, respectively, for a scan based on the 13 actual events and event energies observed by XENON10~\cite{Angle:2009xb} (albeit using our hypothetical experiment).}

\begin{table}[t]
\begin{center}
{\small
\begin{tabular}{p{3.5cm} p{3cm} r}
\toprule
\textbf{Quantity}& \textbf{Benchmark 1}& \multicolumn{1}{c}{\textbf{Benchmark 2}} \\ \toprule
$\mzero$ (GeV)           &  $1849.67$    & $3886.19$ \\
$\mhalf$ (GeV)       &  $563.78$    & $1152.82$ \\
$\azero$ (GeV)          &  $933.13$    & $1529.73$ \\
$\tanb$     &  $54.49$    & $53.09$ \\ \midrule
$\mtpole$ (GeV)      &  $172.50$    & $172.51$ \\
$m_b (m_b)^{\overline{MS}}$ (GeV)& $4.21$  & $4.21$ \\
$\alphas$       &  $0.1186$ & $0.1180$ \\
$1/\alphaemmz$  & $127.929$ & $127.954$ \\ \midrule
$\sigma^{SI}_p$ (pb) & 8.3 $\times 10^{-8}$  & 2.3 $\times 10^{-7}$ \\
$m_{\tilde\chi^0_1}$ (GeV)  &  $235.49$    & $442.74$ \\ \bottomrule
\end{tabular}
} \caption[aa]{\footnotesize{Values of model and nuisance parameters at our benchmark points, as well as the corresponding DD quantities $\sigma^{SI}_p$ and $m_{\tilde\chi^0_1}$.}} \label{tab:params4BMs}
\end{center}
\end{table}

As the next step, we should simulate some pseudo-experiments based on our DD likelihood function (\eq{eqn:DDLike}). The process is as follows: we begin with a benchmark point and fix the true values of our parameters to the corresponding benchmark values. The likelihood then represents probabilities for obtaining particular sets of experimental data for the assumed true values of model parameters. Based on these probabilities, one can generate some random values for the involved random variables (or the data points). The random variables in our case include the number of observed DD events and the measured recoil energies. We simulate 100 pseudo-experiments (with 100 random sets of generated pseudo-data) for each benchmark point. The data is generated by taking a random number of events according to a Poisson distribution with average $\mu$; each of these events is then given a random energy according to the distribution given by \eq{eqn:fE}. We then feed each set of generated data into our scanning machinery as a new set of experimental data and scan over the parameters. The scans are performed over the eight model parameters with ranges given in~\tab{tab:paramranges}. 

\begin{table}[t]
\begin{center}
{\small
\begin{tabular}{p{3.5cm} p{3cm} r}
\toprule
\textbf{Parameter}& \textbf{Lower limit}& \multicolumn{1}{c}{\textbf{Upper limit}} \\ \toprule
$\mzero$ (GeV)           &  $60$    & $4000$ \\
$\mhalf$ (GeV)       &  $60$    & $4000$ \\
$\azero$ (GeV)          &  $-7000$    & $7000$ \\
$\tanb$     &  $2$    & $65$ \\ \midrule
$\mtpole$ (GeV)      &  $163.7$    & $178.1$ \\
$m_b (m_b)^{\overline{MS}}$ (GeV)& $3.92$  & $4.48$ \\
$\alphas$       &  $0.1096$ & $0.1256$ \\
$1/\alphaemmz$  & $127.846$ & $127.990$ \\ \bottomrule
\end{tabular}
} \caption[aa]{\footnotesize{Ranges of model and nuisance parameters employed in our scans.}} \label{tab:paramranges}
\end{center}
\end{table}

In order to ensure that our sample points are physically self-consistent, we assign extremely small (almost zero) values to the likelihood for the unphysical points. These points are the ones for which no self-consistent solutions to the RGEs exist, EWSB conditions are not fulfilled, some particle masses are tachyonic, or the neutralino is not the LSP. In addition to the DD data and physicality conditions, we also apply experimental constraints on the measurements of the SM nuisance parameters. In \textsf{SuperBayeS}, the likelihoods for these quantities are modelled by Gaussian functions; the mean values and standard deviations we use in our analysis are given in~\tab{tab:Nuisances}.

\begin{table}[t]
\begin{center}
{\small
\begin{tabular}{p{3cm} p{3cm} p{4cm} r} \toprule
\textbf{Observable} &   \textbf{Mean value} & \textbf{Standard deviation} & \textbf{Reference} \\
\toprule
$\mtpole$ (GeV)           &  $172.6$    & $1.4$& \cite{topmass:mar08} \\
$m_b (m_b)^{\overline{MS}}$ (GeV) & $4.20$  & $0.07$ & \cite{pdg07} \\
$\alphas$       &   $0.1176$   & $0.002$ & \cite{pdg07}\\
$1/\alphaemmz$  & $127.955$ & $0.03$ & \cite{Hagiwara:2006jt}
\\ \bottomrule
\end{tabular}
} \caption[aa]{\footnotesize{Measurements of SM nuisance parameters used in the analysis.}} \label{tab:Nuisances}
\end{center}
\end{table}

For each scan, we calculate 1D marginal posteriors and profile likelihoods for different model parameters and derived observables. Finally, for every parameter or observable and by looking at all the 100 scans, we count the number of times the corresponding benchmark value falls into a certain credible or confidence interval. This gives the approximate degree of coverage for the corresponding level of confidence. We perform this process for both our benchmark points and evaluate the coverage for $68.3\%$ ($1\sigma$) and $95.4\%$ ($2\sigma$) confidence levels.

For the scans over the parameter space, we employ both flat and logarithmic (log) priors. A log prior for a parameter $\theta_i$ (which receives only positive values) is defined as uniform in $\ln\theta_i$. The main motivation for choosing log priors is that they represent prior `ignorance' instead of prior knowledge. The use of these `non-informative' priors is justified in attempts for reducing the subjectivity of Bayesian interval construction, although the actual power of Bayesian statistics is in that it provides the possibility of including prior knowledge rather than ignorance~\cite{Feldman:1998}. The other motivation for log priors in terms of the sampling algorithm is that samples based on uniform priors are not distributed uniformly in a hypersphere in the parameter space, which one usually requires in order to correctly map a multi-dimensional function (such as our likelihood). Log priors are believed to alleviate this problem (for an example where log priors are shown to improve the statistical coverage, see Ref.~\cite{Heinrich:2006}).

\section{Results and discussion} \label{sec:results}
\subsection{One-dimensional confidence intervals and coverage evaluations} \label{sec:detect}

\begin{table}[t]
\begin{center}
\begin{tabular}{|l | >{\centering\arraybackslash}p{1.2cm} | >{\centering\arraybackslash}p{1cm} | >{\centering\arraybackslash}p{1cm} | >{\centering\arraybackslash}p{1cm} | >{\centering\arraybackslash}p{1cm} | >{\centering\arraybackslash}p{1cm} | >{\centering\arraybackslash}p{1cm} | >{\centering\arraybackslash}p{1cm} | >{\centering\arraybackslash}p{1cm}|} \cline{3-10}
\multicolumn{2}{c|}{}& \multicolumn{4}{c|}{\textbf{Benchmark 1}} & \multicolumn{4}{c|}{\textbf{Benchmark 2}} \bigstrut[t]\\ \cline{3-10}
\multicolumn{2}{c|}{}& \multicolumn{2}{c|}{\textbf{Conf. int.}} & \multicolumn{2}{c|}{\textbf{Cred. int.}} & \multicolumn{2}{c|}{\textbf{Conf. int.}} & \multicolumn{2}{c|}{\textbf{Cred. int.}} \bigstrut\\ \cline{3-10}
\multicolumn{2}{c|}{} & $1\sigma$ & $2\sigma$ & $1\sigma$ & $2\sigma$ & $1\sigma$ & $2\sigma$ & $1\sigma$ & $2\sigma$ \bigstrut\\ \hhline{--========}
\multirow{8}{*}{\begin{sideways}\textbf{Flat priors}\end{sideways}} & $\mzero$           &  \textcolor{black}{\textbf{96}} & \textcolor{black}{\textbf{100}} & \textcolor{black}{\textbf{98}} & \textcolor{black}{\textbf{100}} & \textcolor{red}{\textbf{\textbf{62}}} & \textcolor{red}{\textbf{\textbf{93}}} & \textcolor{red}{\textbf{\textbf{0}}} & \textcolor{red}{\textbf{\textbf{1}}} \bigstrut \\ \cline{2-10}\cline{2-10}
 & $\mhalf$           &  \textcolor{black}{\textbf{78}} & \textcolor{green}{\textbf{97}} & \textcolor{red}{\textbf{45}} & \textcolor{green}{\textbf{97}} & \textcolor{red}{\textbf{39}} & \textcolor{red}{\textbf{89}} & \textcolor{red}{\textbf{0}} & \textcolor{red}{\textbf{0}} \bigstrut\\ \cline{2-10}
 & $\azero$           &  \textcolor{black}{\textbf{100}} & \textcolor{black}{\textbf{100}} & \textcolor{black}{\textbf{91}} & \textcolor{black}{\textbf{100}} & \textcolor{black}{\textbf{96}} & \textcolor{black}{\textbf{100}} & \textcolor{red}{\textbf{0}} & \textcolor{green}{\textbf{95}} \bigstrut\\ \cline{2-10}
 & $\tanb$           &  \textcolor{black}{\textbf{82}} & \textcolor{black}{\textbf{100}} & \textcolor{red}{\textbf{19}} & \textcolor{black}{\textbf{100}} & \textcolor{black}{\textbf{99}} & \textcolor{black}{\textbf{100}} & \textcolor{black}{\textbf{75}} & \textcolor{black}{\textbf{100}} \bigstrut\\ \cline{2-10}
 & $m_{\tilde\chi^0_1}$           &  \textcolor{black}{\textbf{75}} & \textcolor{green}{\textbf{97}} & \textcolor{red}{\textbf{18}} & \textcolor{green}{\textbf{97}} & \textcolor{red}{\textbf{45}} & \textcolor{red}{\textbf{93}} & \textcolor{red}{\textbf{0}} & \textcolor{red}{\textbf{0}} \bigstrut\\\cline{2-10}
 & $\sigma^{SI}_p$           &  \textcolor{black}{\textbf{76}} & \textcolor{black}{\textbf{98}} & \textcolor{red}{\textbf{53}} & \textcolor{green}{\textbf{96}} & \textcolor{red}{\textbf{51}} & \textcolor{red}{\textbf{87}} & \textcolor{red}{\textbf{0}} & \textcolor{red}{\textbf{22}} \bigstrut\\ 
\hline\hline
\multirow{8}{*}{\begin{sideways}\textbf{Log priors}\end{sideways}} & $\mzero$           &  \textcolor{black}{\textbf{96}} & \textcolor{black}{\textbf{100}} & \textcolor{red}{\textbf{15}} & \textcolor{green}{\textbf{94}} & \textcolor{red}{\textbf{17}} & \textcolor{red}{\textbf{47}} & \textcolor{red}{\textbf{0}} & \textcolor{red}{\textbf{0}} \bigstrut\\ \cline{2-10}
 & $\mhalf$           &  \textcolor{green}{\textbf{67}} & \textcolor{red}{\textbf{92}} & \textcolor{red}{\textbf{2}} & \textcolor{red}{\textbf{30}} & \textcolor{red}{\textbf{1}} & \textcolor{red}{\textbf{17}} & \textcolor{red}{\textbf{0}} & \textcolor{red}{\textbf{0}} \bigstrut\\ \cline{2-10}
 & $\azero$           &  \textcolor{black}{\textbf{99}} & \textcolor{black}{\textbf{100}} & \textcolor{red}{\textbf{43}} & \textcolor{red}{\textbf{91}} & \textcolor{black}{\textbf{91}} & \textcolor{black}{\textbf{100}} & \textcolor{red}{\textbf{0}} & \textcolor{red}{\textbf{24}} \bigstrut\\ \cline{2-10}
 & $\tanb$           &  \textcolor{black}{\textbf{93}} & \textcolor{black}{\textbf{100}} & \textcolor{red}{\textbf{16}} & \textcolor{red}{\textbf{91}} & \textcolor{black}{\textbf{99}} & \textcolor{black}{\textbf{100}} & \textcolor{red}{\textbf{38}} & \textcolor{black}{\textbf{99}} \bigstrut\\ \cline{2-10}
 & $m_{\tilde\chi^0_1}$           &  \textcolor{red}{\textbf{57}} & \textcolor{red}{\textbf{88}} & \textcolor{red}{\textbf{17}} & \textcolor{red}{\textbf{37}} & \textcolor{red}{\textbf{2}} & \textcolor{red}{\textbf{15}} & \textcolor{red}{\textbf{23}} & \textcolor{red}{\textbf{23}} \bigstrut\\ \cline{2-10}
 & $\sigma^{SI}_p$           &  \textcolor{green}{\textbf{71}} & \textcolor{green}{\textbf{97}} & \textcolor{red}{\textbf{22}} & \textcolor{red}{\textbf{65}} & \textcolor{red}{\textbf{15}} & \textcolor{red}{\textbf{59}} & \textcolor{red}{\textbf{0}} & \textcolor{red}{\textbf{1}} \bigstrut\\ \hline
\end{tabular}
\caption[aa]{\footnotesize{Results of our coverage evaluations. Each entry shows the number of times the hypothesised true value of a parameter or derived quantity falls into $68.3 \%$ ($1\sigma$) or $95.4 \%$ ($2\sigma$) frequentist confidence or Bayesian credible intervals for 100 pseudo-experiments simulated based on the likelihood. The numbers are given for the two benchmark points we have chosen in our analysis and for both flat and log-prior scans. Numbers in red and black show under- and over-coverage, respectively, while the ones in green show correct coverage. Green numbers are the ones that fall within $[64,72]$ and $[94,97]$ intervals for $1\sigma$ and $2\sigma$ confidence levels, respectively. A binomial estimate of errors has been used to determine whether the coverage is correct in each case.}} \label{tab:coverage}
\end{center}
\end{table}

We summarise our results in \tab{tab:coverage}. For each benchmark point and specific choice of priors, we list the number of times that the corresponding benchmark values for the CMSSM parameters $\mzero$, $\mhalf$, $\azero$ and $\tanb$ fall within their 1D $68.3 \%$ ($1\sigma$) or $95.4 \%$ ($2\sigma$) confidence intervals. We also add the two derived quantities $m_{\tilde\chi^0_1}$ and $\sigma^{SI}_p$ to the list because those are the main quantities upon which the generated data depend. Since we use 100 realisations for each pseudo-true benchmark point, the presented numbers span a range between 0 and 100. For comparison, we also give the coverage for Bayesian credible intervals, although we do not expect proper coverage in those cases. Numbers in red and black show under- and over-coverage, respectively, while numbers in green indicate correct coverage (to within errors).

In order to verify whether a particular number in the table corresponds to under-, over- or appropriate coverage, we use binomial errors on the trial counts for the considered confidence levels $68.3 \%$ and $95.4 \%$. The errors are calculated as the square root of the expression 
\begin{equation} \label{eqn:binomial}
  \sigma^2=np(1-p)
\end{equation}
for the variance of a binomial distribution.  Here $n$ is the number of trials, which is $100$ in our case, and the result of each trial is assumed to be `success' (where the confidence interval in question contains the pseudo-true value) with probability $p$, and `failure' with probability $1-p$. In our cases $p$ is $0.683$ and $0.954$, for $68.3 \%$ ($1\sigma$) and $95.4 \%$ ($2\sigma$) confidence levels, respectively. This means that a count of successes for a $1\sigma$ ($2\sigma$) confidence level shows exact coverage if it lies in the range $[64,72]$ ($[94,97]$); it is then coloured green in~\tab{tab:coverage}. Under- and over-coverage are defined in a similar way for numbers outside the corresponding ranges for the reconstructed $1\sigma$ or $2\sigma$ intervals, $[64,72]$ or $[94,97]$, respectively (coloured red and black in~\tab{tab:coverage}).   

We should point out that our estimation here based on a binomial process is in fact an estimate of the statistical error that comes from having a finite number of pseudo-experiments to estimate the coverage with. In a similar coverage study for the CMSSM, based on an ATLAS likelihood~\cite{Bridges2010}, errors are estimated by first working out how much realisation noise is produced by~\MN~and the neural networks employed for accelerating the scans in that study. This provides the standard deviation of the upper and lower ends of the one-dimensional confidence 
intervals on the CMSSM parameters when doing repeated scans for the same set of pseudo-data.  All the derived confidence intervals (i.e. from every pseudo-experiment) are then shifted uniformly up or down by the standard deviation on the upper or lower limit, and error bars assigned to coverages based on the differences induced in the derived coverage values. This estimate provides a measure of the random bias (i.e. systematic error) induced by realisation noise in the scans, due to a finite number of scans (one) for each pseudo-experiment.

In principle one should estimate both errors (the statistical, as in our calculations, \textit{and} the systematic, as in Ref.~\cite{Bridges2010}), and include both in the total error budget on coverages.  On the basis of the error bars in Ref.~\cite{Bridges2010} however, it seems that the possible sampling bias estimated in that analysis is usually smaller than the statistical error included here.  The statistical error we calculate should thus mostly dominate in our case, so we are well justified in using only the binomial error in this study.  Given the larger number of pseudo-experiments in Ref.~\cite{Bridges2010} on the other hand, sampling noise (i.e. systematics) should dominate the error budget in that analysis, so the binomial error can be safely neglected.

\begin{figure}[t]
\begin{center}
\subfigure{\includegraphics[width=0.37\linewidth, trim = 45 230 55 120, clip=true]{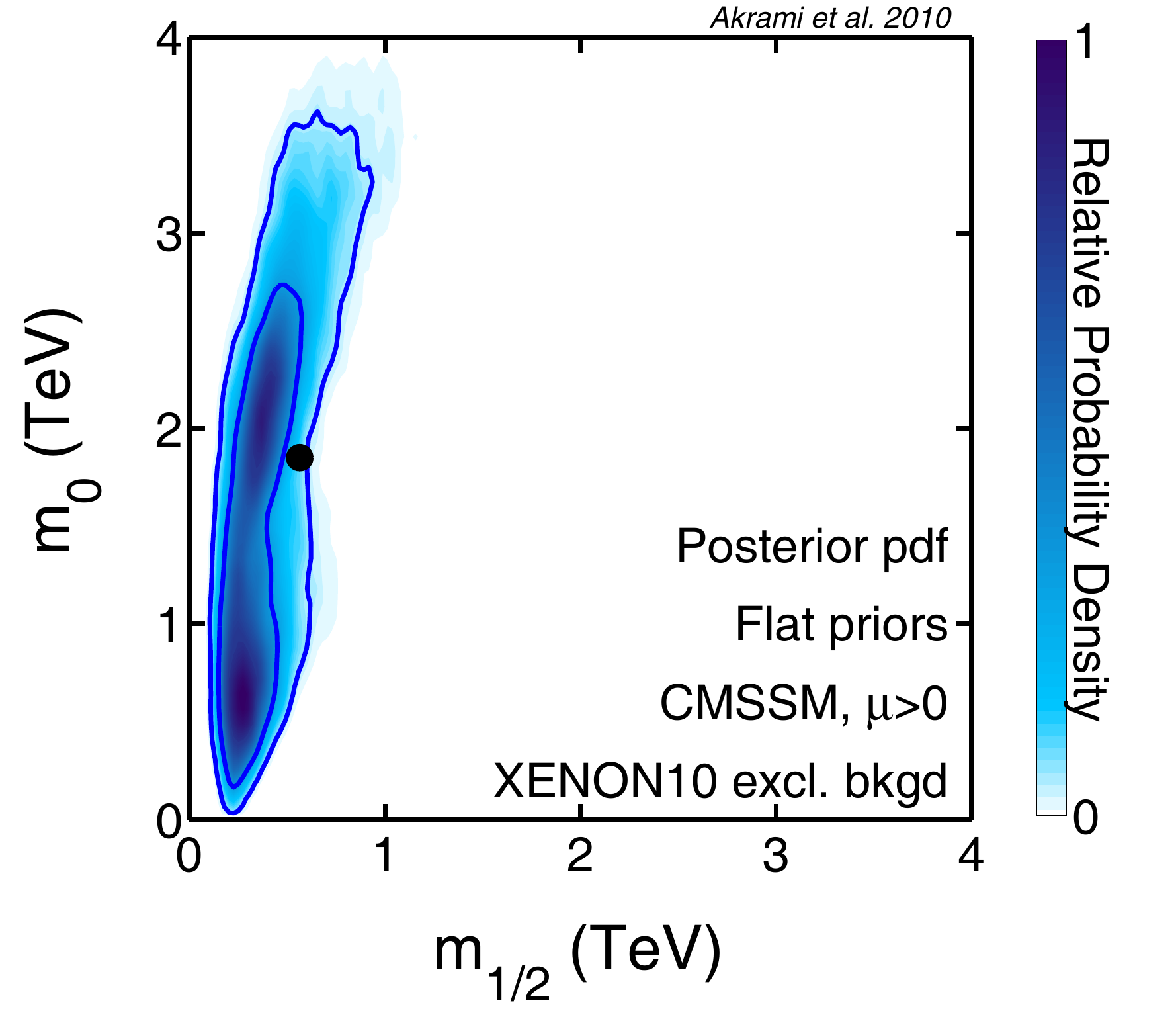}}
\subfigure{\includegraphics[width=0.37\linewidth, trim = 45 230 55 120, clip=true]{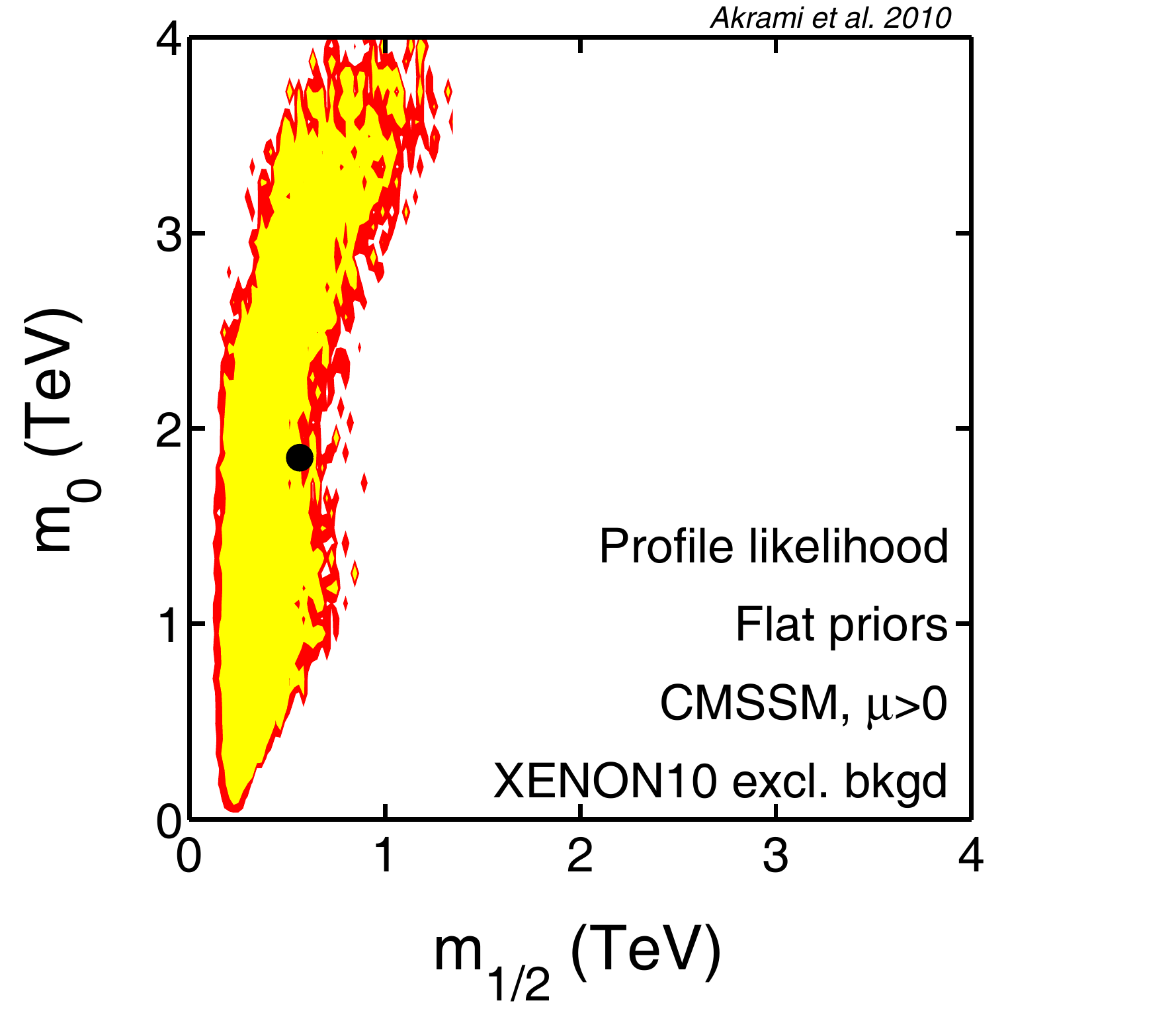}}\\
\subfigure{\includegraphics[width=0.37\linewidth, trim = 45 230 55 120, clip=true]{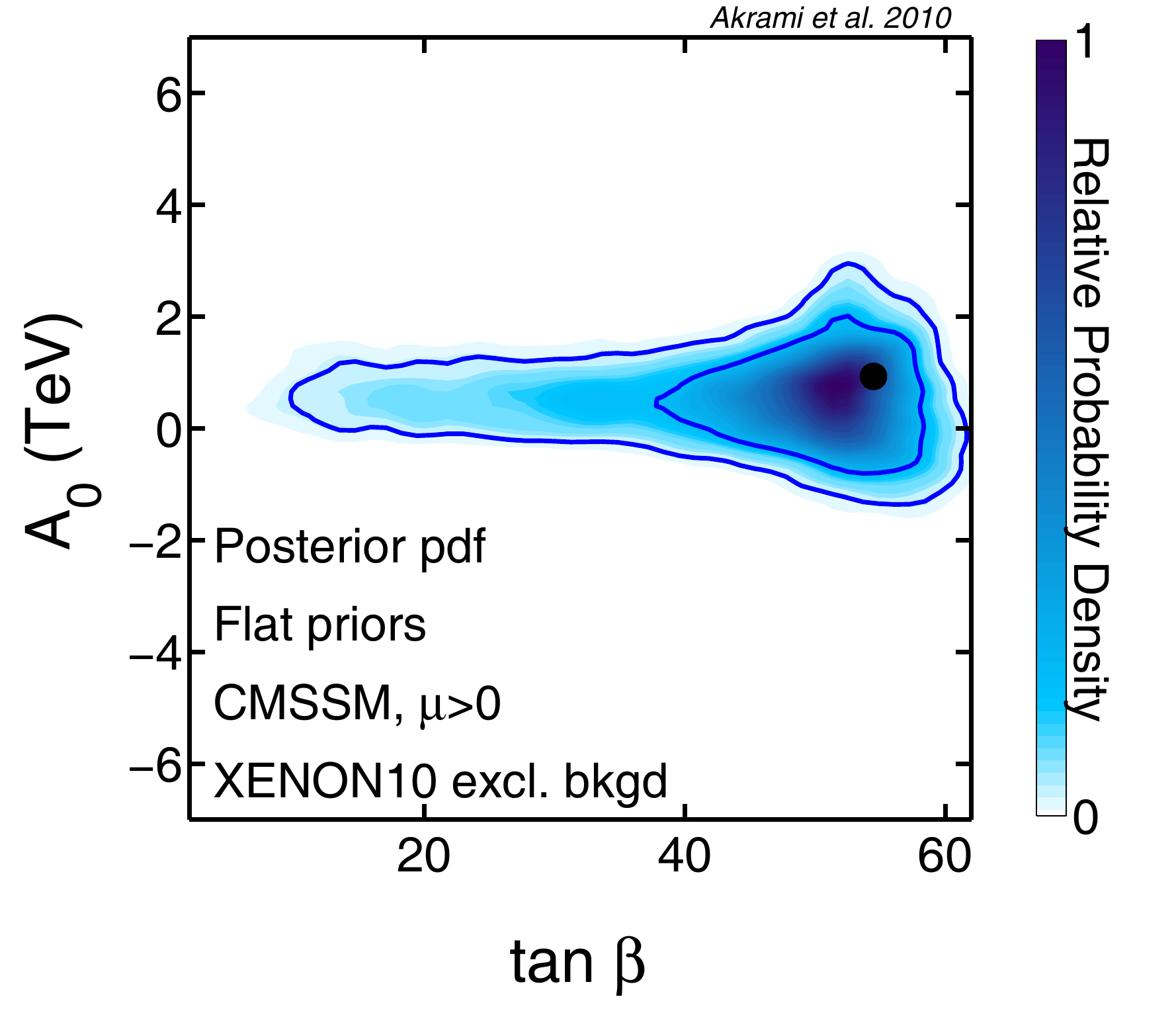}}
\subfigure{\includegraphics[width=0.37\linewidth, trim = 45 230 55 120, clip=true]{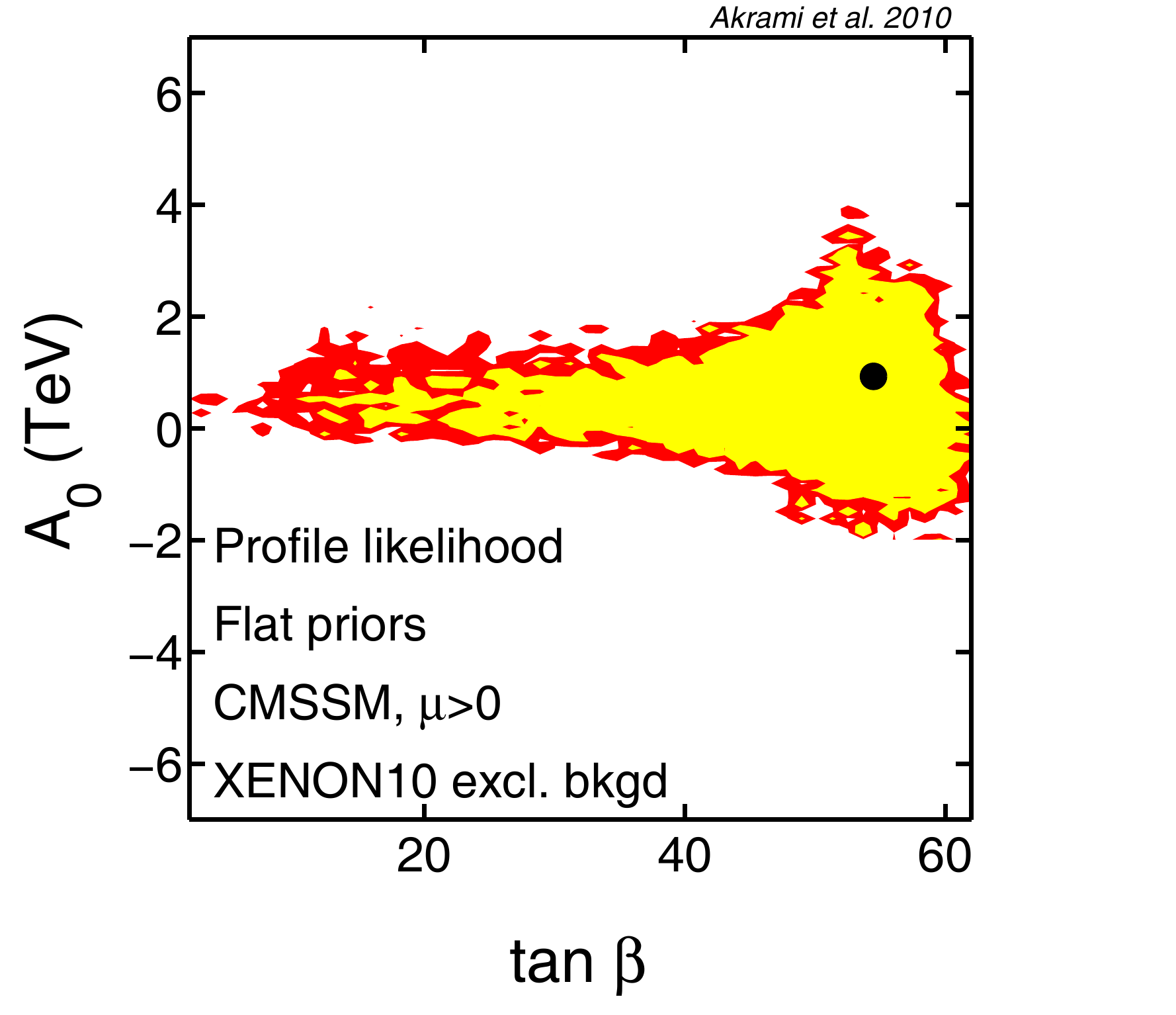}}\\
\subfigure{\includegraphics[width=0.37\linewidth, trim = 45 230 55 120, clip=true]{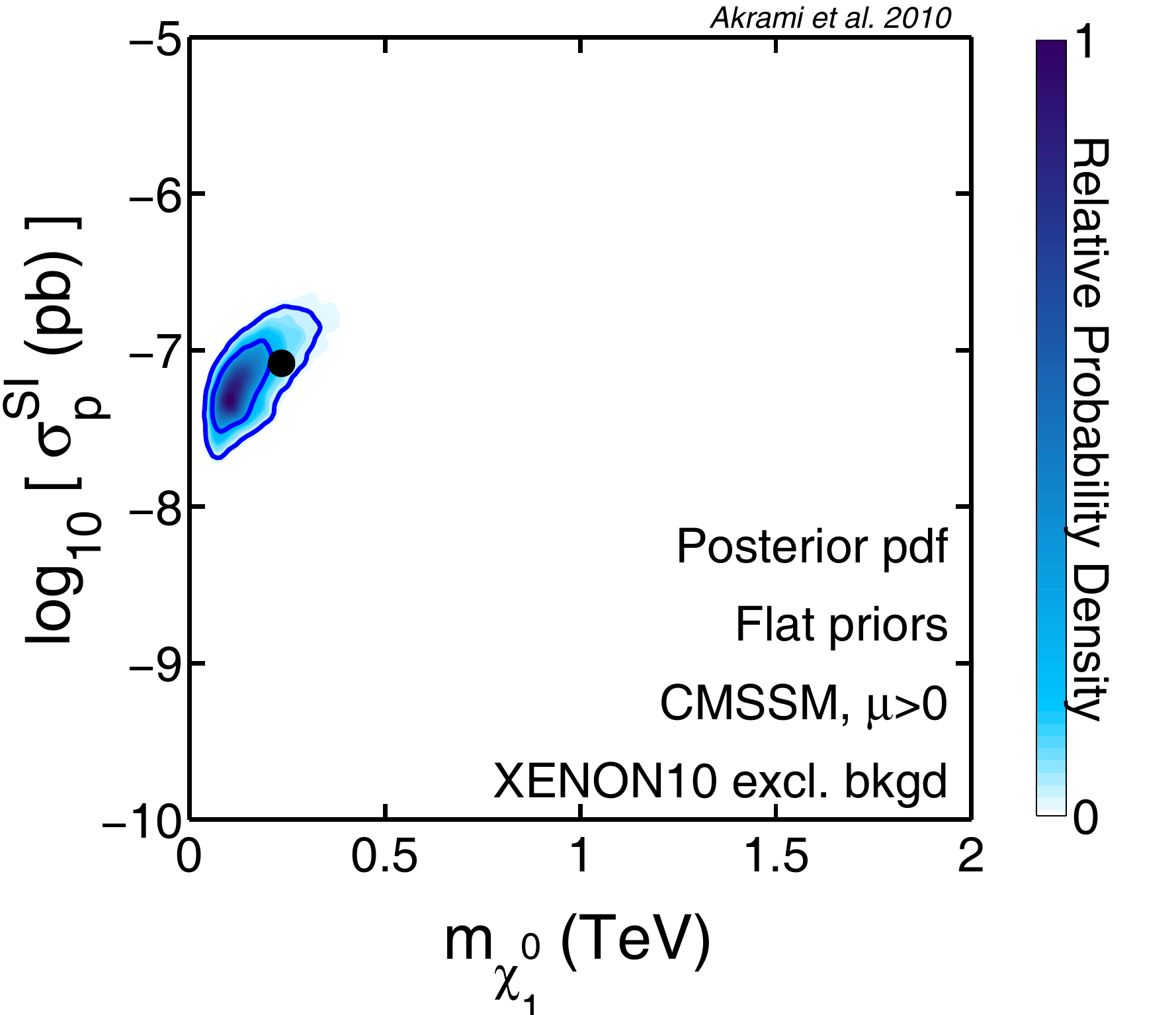}}
\subfigure{\includegraphics[width=0.37\linewidth, trim = 45 230 55 120, clip=true]{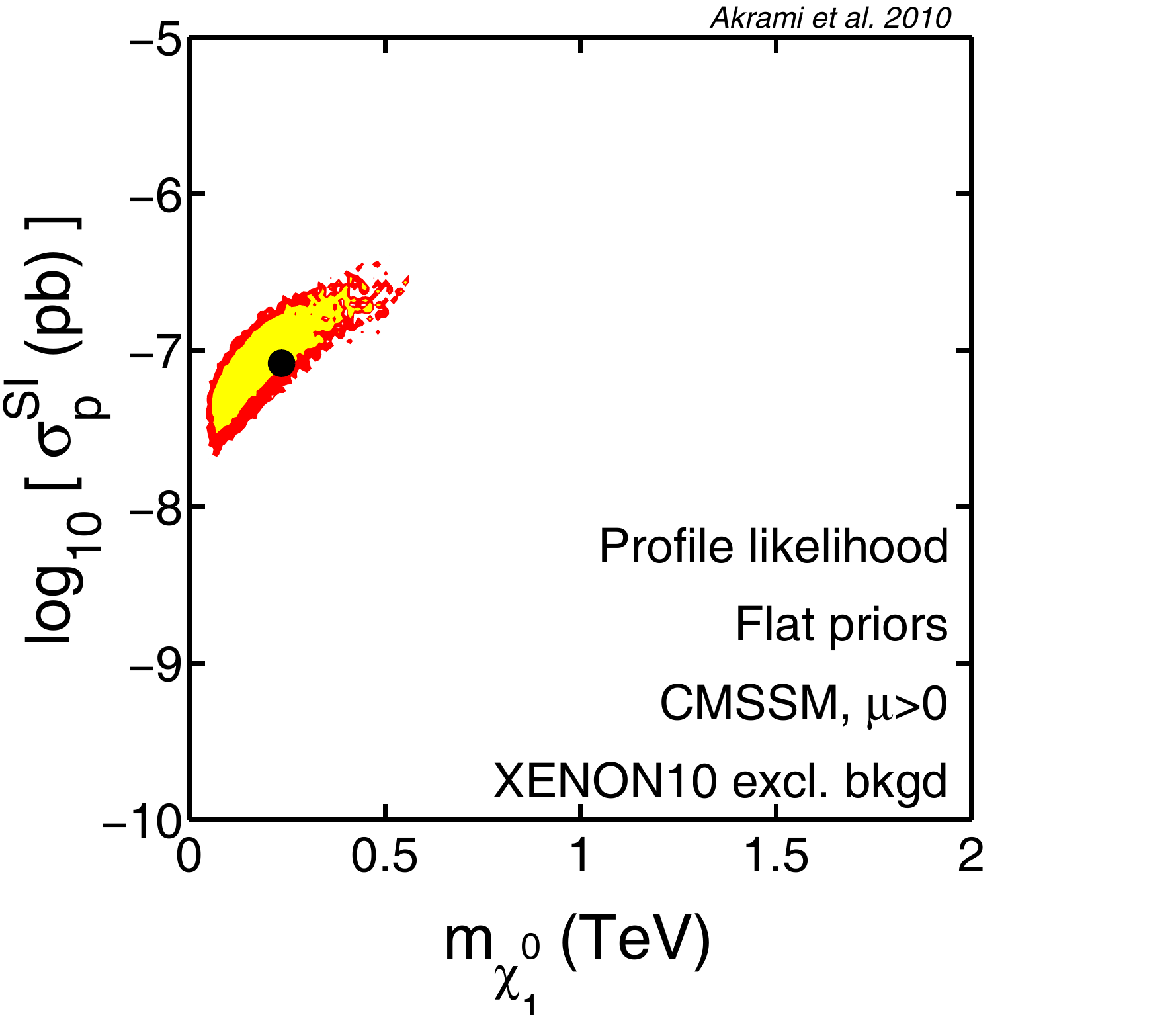}}\\
\caption[aa]{\footnotesize{Two-dimensional marginal posteriors and profile likelihoods for one typical coverage scan over the CMSSM parameter space using a set of random DD data generated from the hypothetical XENON10 likelihood. In this case, pseudo-true parameter values are set to those of benchmark 1. Plots are shown in $m_0$-$m_{1/2}$ and $A_0$-$\tanb$ planes, as well as for the spin-independent scattering cross-section of the neutralino and a proton $\sigma^{SI}_p$ versus the neutralino mass $m_{\tilde\chi^0_1}$. The inner and outer contours in each panel represent $68.3\%$ ($1\sigma$) and $95.4\%$ ($2\sigma$) confidence levels, respectively. Black dots show the benchmark values.}}\label{fig:2D_benchmark1}
\end{center}
\end{figure}

\begin{figure}[t]
\begin{center}
\subfigure{\includegraphics[width=0.37\linewidth, trim = 45 230 55 120, clip=true]{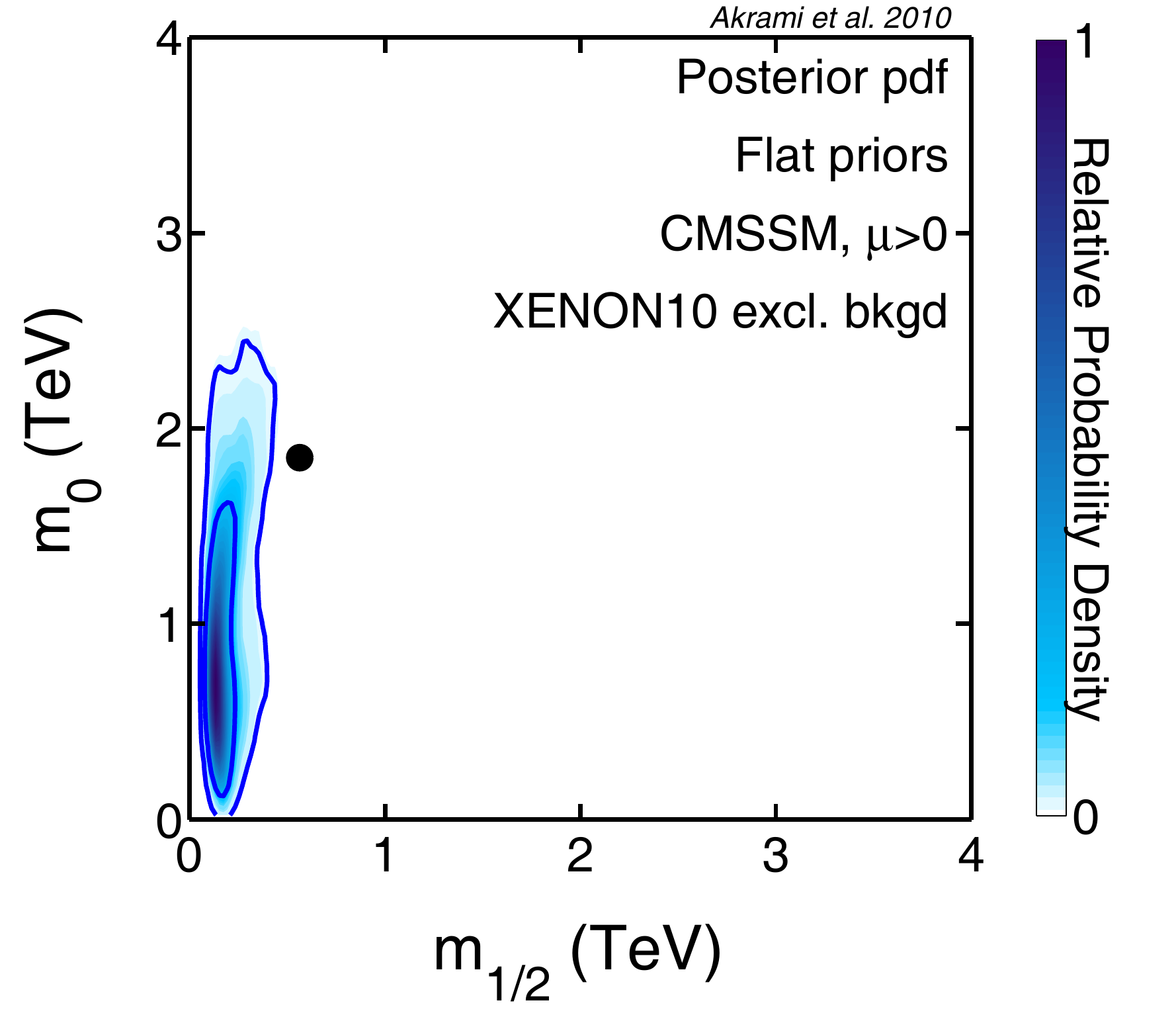}}
\subfigure{\includegraphics[width=0.37\linewidth, trim = 45 230 55 120, clip=true]{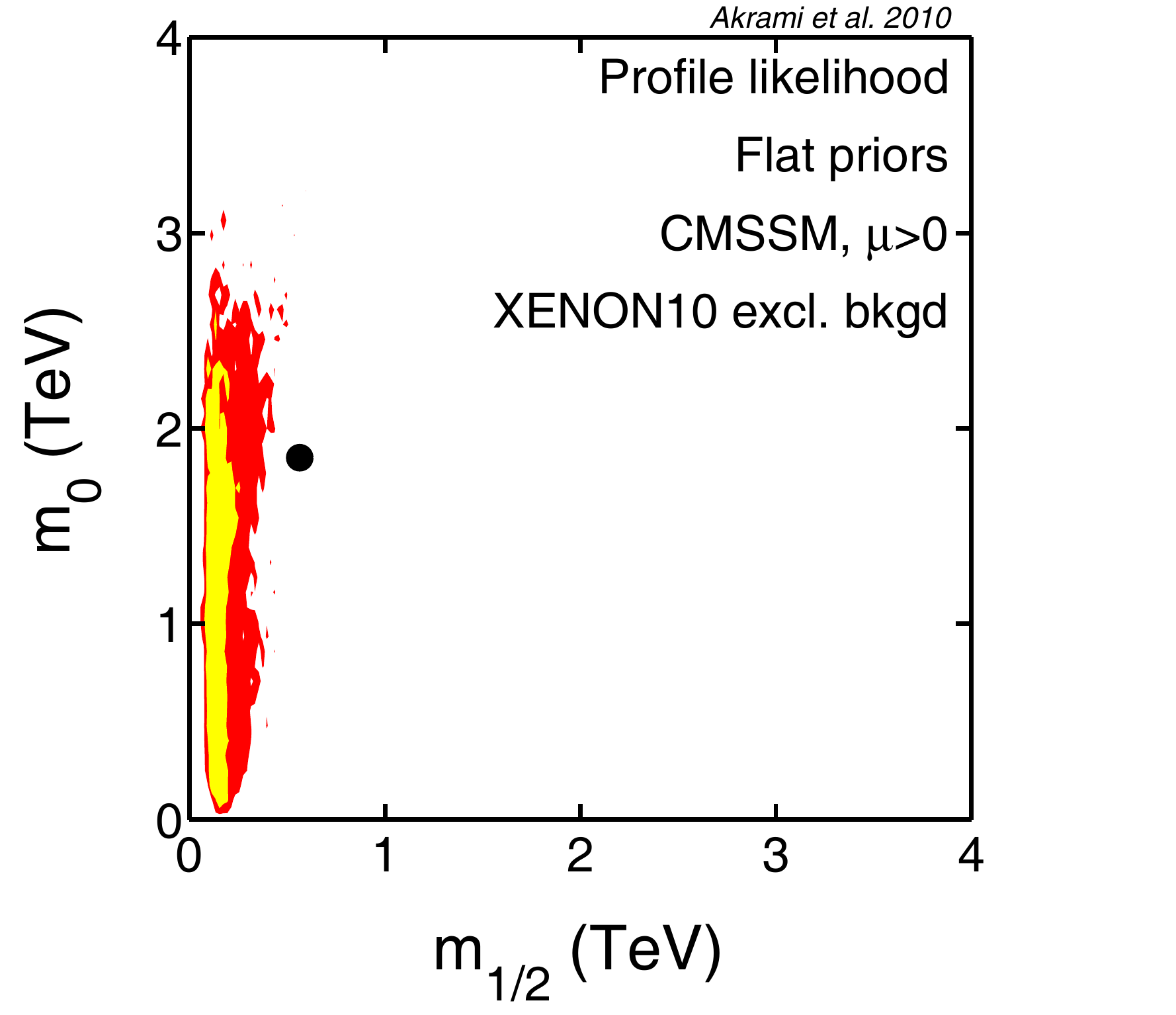}}\\
\subfigure{\includegraphics[width=0.37\linewidth, trim = 45 230 55 120, clip=true]{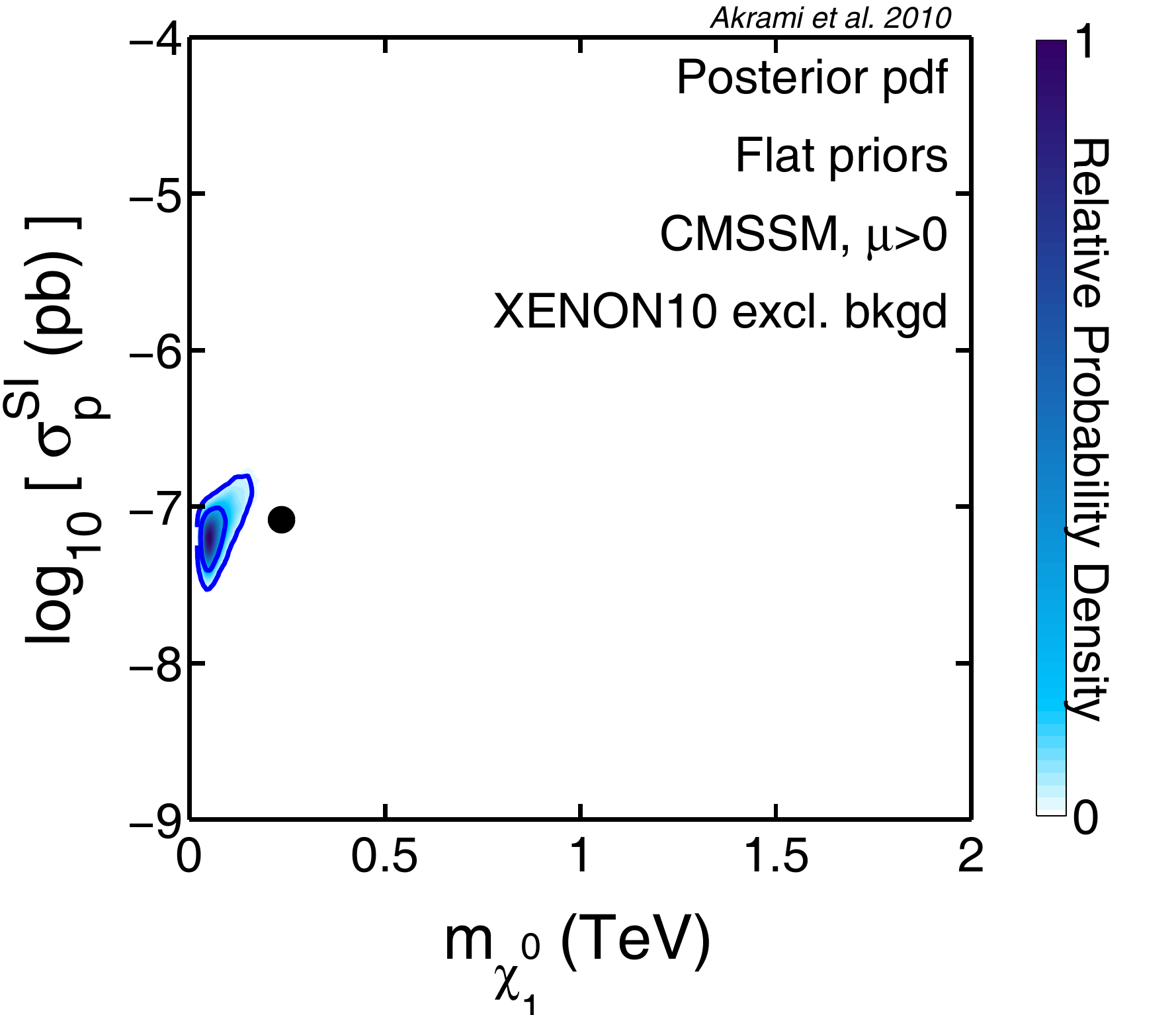}}
\subfigure{\includegraphics[width=0.37\linewidth, trim = 45 230 55 120, clip=true]{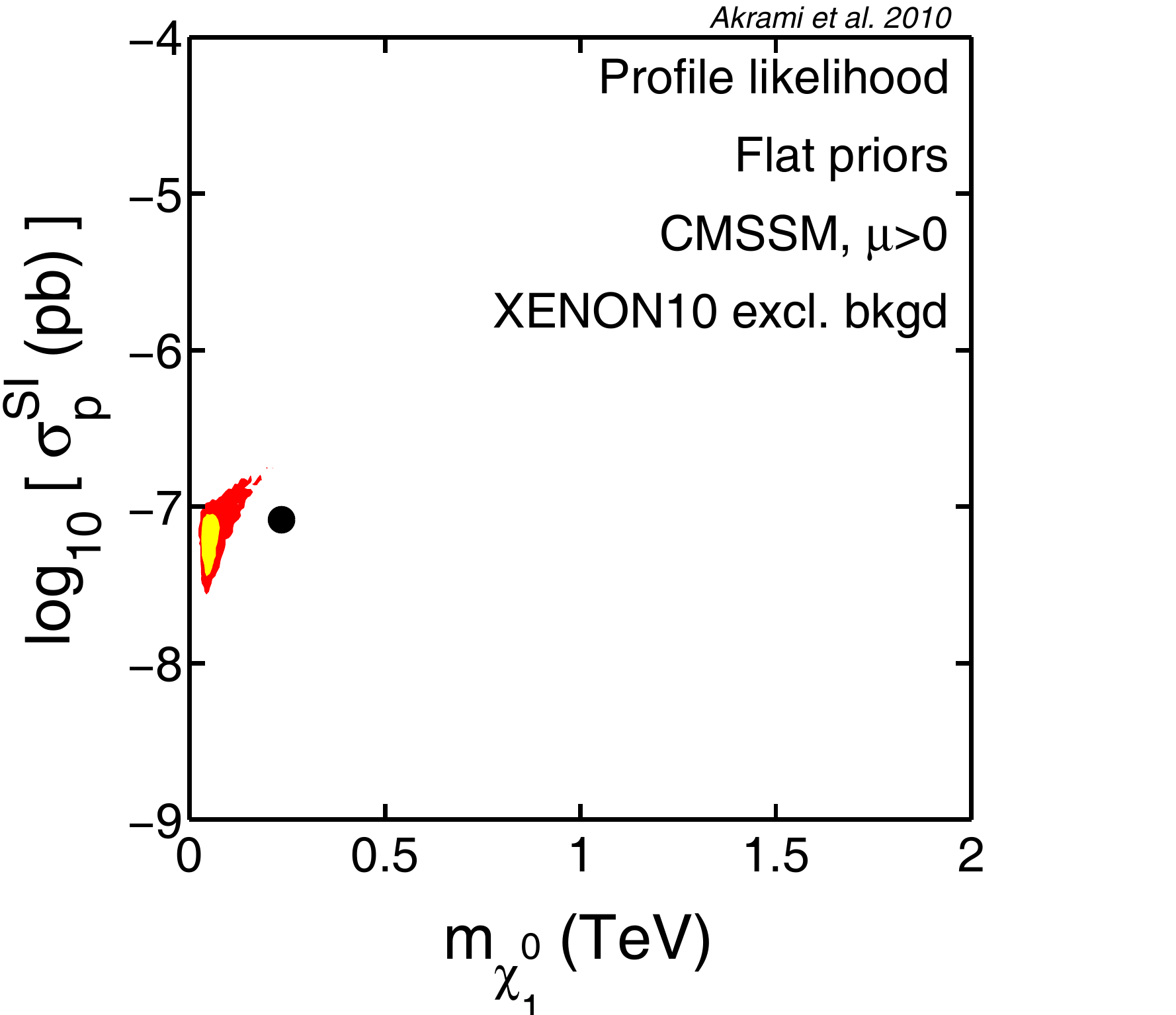}}\\
\caption[aa]{\footnotesize{As in~\fig{fig:2D_benchmark1}, but for a scan with $1\sigma$ and $2\sigma$ credible and confidence regions that do not contain the true parameters. In cases with proper coverage these should happen in about $32\%$ and $5\%$ of the times for $1\sigma$ and $2\sigma$ regions, respectively. For cases with significant under-coverage (as exhibited in some of our results), these types of plots occur much more frequently}. Again, pseudo-true parameter values are set to those of benchmark 1.}\label{fig:2D_benchmark1_2}
\end{center}
\end{figure}

Before we discuss the results given in~\tab{tab:coverage}, we show in~\fig{fig:2D_benchmark1} how the outcomes of a typical scan using our generated data may look. The figure shows the obtained two-dimensional (2D) credible and confidence regions for Bayesian marginal posterior PDFs and frequentist profile likelihoods, respectively, for a scan with benchmark 1 representing the true parameter values, and flat priors imposed on $\mzero$ and $\mhalf$. Plots are given for the four CMSSM parameters $\mzero$, $\mhalf$, $\azero$ and $\tanb$, and the two derived quantities $\sigma^{SI}_p$ and $m_{\tilde\chi^0_1}$. The $\sigma^{SI}_p$-$m_{\tilde\chi^0_1}$ planes are particularly interesting for DD experiments. The inner and outer contours in each panel (in dark and light blue for marginal posteriors and in yellow and red for profile likelihoods) represent $68.3\%$ ($1\sigma$) and $95.4\%$ ($2\sigma$) confidence levels, respectively. Black dots indicate the benchmark values corresponding to benchmark 1. We also show in~\fig{fig:2D_benchmark1_2} the outcomes of one of the scans for which the true values are located outside the $1\sigma$ and $2\sigma$ credible and confidence regions. For cases where the coverage requirement is appropriately fulfilled, these types of plots should appear in $31.7\%$ and $4.6\%$ of the scans for $1\sigma$ and $2\sigma$ regions, respectively. For cases with significant under-coverage, these occur in larger fractions of times. \fig{fig:2D_benchmark1} and~\fig{fig:2D_benchmark1_2} also indicate how the inferred confidence regions may vary in different scans with different realisations of the experiment.

Let us now look at~\tab{tab:coverage}. The two first columns, corresponding to the confidence intervals for benchmark 1 are dominated by over-coverage, especially for flat-prior scans. The under-coverage for log priors, which happens only for $\mhalf$ and $m_{\tilde\chi^0_1}$, is not severe. In particular for $\mhalf$ in this case, the numbers are very close to the exact-coverage values and are likely to tend to those in the large number limit (i.e. when the number of pseudo-experiments tends to infinity).

For benchmark 2, on the other hand, our results show severe under-coverage; only for $\azero$ and $\tanb$ do the intervals over-cover the true parameters. By comparing the flat- and log-prior cases for benchmark 2, we observe that for cases with coverage less than the desired value, this under-coverage deteriorates significantly from flat to log priors, in particular for $\mhalf$ and $m_{\tilde\chi^0_1}$.

In general, for $\mzero$, $\mhalf$, $m_{\tilde\chi^0_1}$ and $\sigma^{SI}_p$, by going from benchmark 1 to benchmark 2, the coverage for frequentist confidence intervals always reduces. Our results also show that the coverage does not change remarkably for $\azero$ and $\tanb$ by changing the benchmarks or priors; intervals always (strongly) over-cover the parameters. 

Looking at the number of times the pseudo-true values are contained within Bayesian credible intervals (and we do this only for comparison), analogous patterns can be recognised: overall, the degree of coverage decreases when going from benchmark 1 to benchmark 2, or from flat priors to log priors. Exceptions are, for example, both $1\sigma$ and $2\sigma$ intervals for $m_{\tilde\chi^0_1}$, where the coverage for benchmark 2 improves by changing priors from flat to log, and for $\tanb$, where the same happens by going from benchmark 1 to benchmark 2 in both flat- and log-prior cases.

Finally, a general comparison of the numbers for frequentist confidence and Bayesian credible intervals indicates that the former offer a substantially better coverage than the latter. This observation is not surprising though because, as we stated earlier, the statistical coverage is a property that is required for frequentist statistics and one in general does not expect a proper coverage for Bayesian inference.

We think that many of the features in our results come from some `sampling effects' that are caused by the explicit and implicit priors imposed on different parameters and observables, as well as the complex structure of the parameter space induced by various physicality constraints. These priors and physicality requirements affect the sampling of the parameter space and consequently the mapping of the likelihood function. In the next section, we describe these sampling effects and explain some of the patterns we see in our results based on them.

\subsection{Sampling effects} \label{sec:confreg}

Before investigating effects of new data on the statistical inference for a model, one needs to examine how well one's scans sample different regions of the parameter space, independent of any information given by the data. For Bayesian statistics, this is obviously an important step because this shows what the effects of priors (and physicality conditions) are on the final inference, i.e. how they impact posterior PDFs when data are added. This makes it possible to know how the data further constrain the parameter space (if at all).

Since no scanning algorithm is flawless, and in most cases (including our implementation) they proceed according to the posterior PDF, these `effective priors' (i.e. the effects from the explicit priors and the physicality conditions) also influence the ability of an algorithm to map the likelihood function and construct frequentist confidence intervals. The reason is that such algorithms tend to sample some regions more than the others, and for relatively strong prior effects this can result in poor or even no mapping of some important parts of the parameter space. These effects can therefore dramatically change the entire statistical inference. Any improper construction of confidence intervals in these cases can consequently induce significant deficiencies in the statistical coverage which are no more than artefacts of the scanning algorithm.  In particular, such biases are independent of the precision of the actual approximate technique used for constructing the frequentist confidence intervals. Although in general we do not expect exact coverage from the profile likelihood method, we do believe that many of the features we observe in our results arise predominantly from these sampling artefacts.

The ideal way to analyse such effects on the coverage is to look at the multi-dimensional distribution of the total sample points generated by priors and physicality conditions only, and then compare the algorithm's tendencies in sampling different regions, in particular regions around the benchmark points selected for the coverage study. This is clearly impossible to display for a parameter space with more than three dimensions.

\begin{figure}[t]
\centering
\subfigure{\includegraphics[scale=0.45, trim = 40 260 100 130, clip=true]{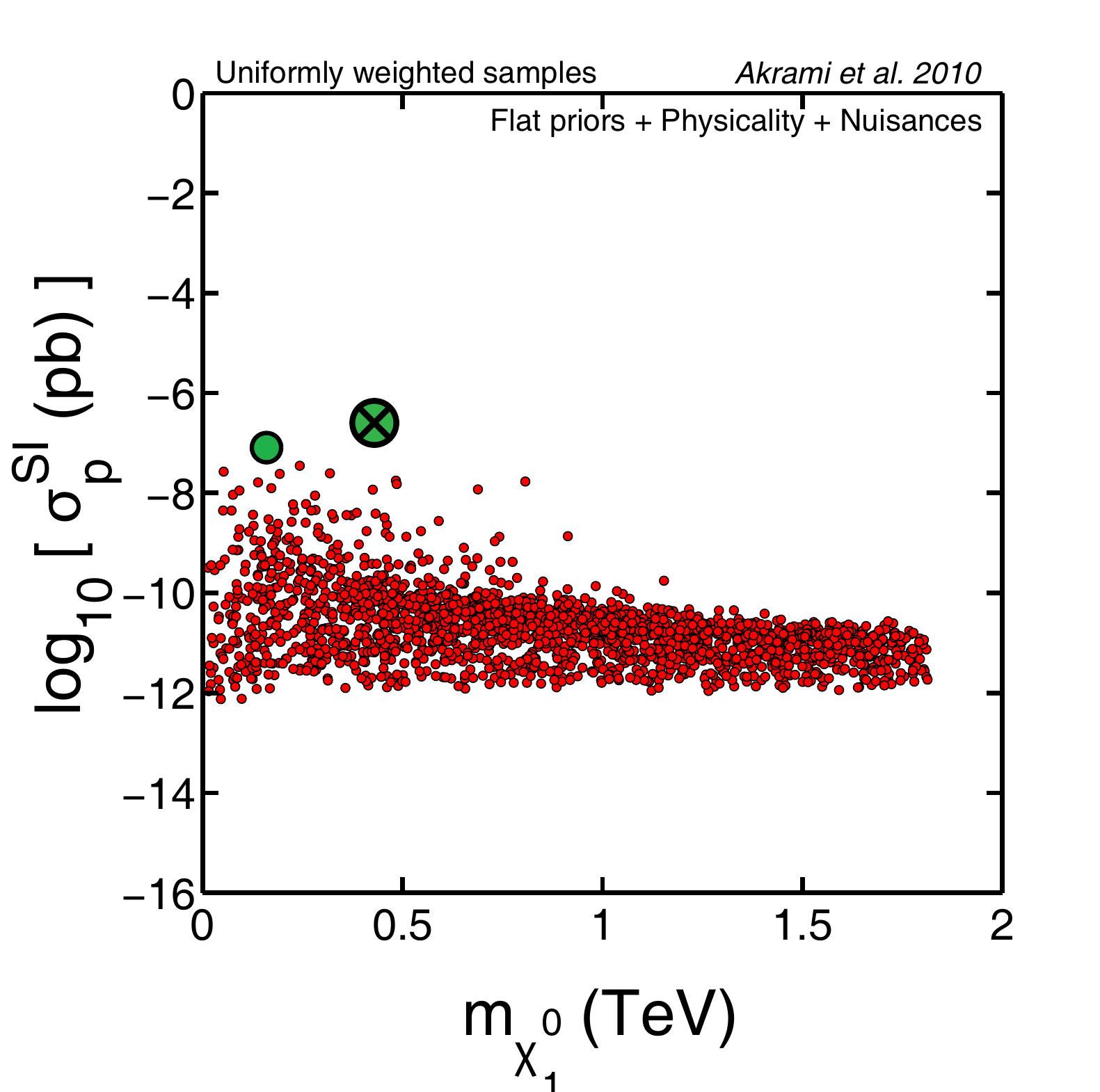}}
\subfigure{\includegraphics[scale=0.45, trim = 40 260 100 130, clip=true]{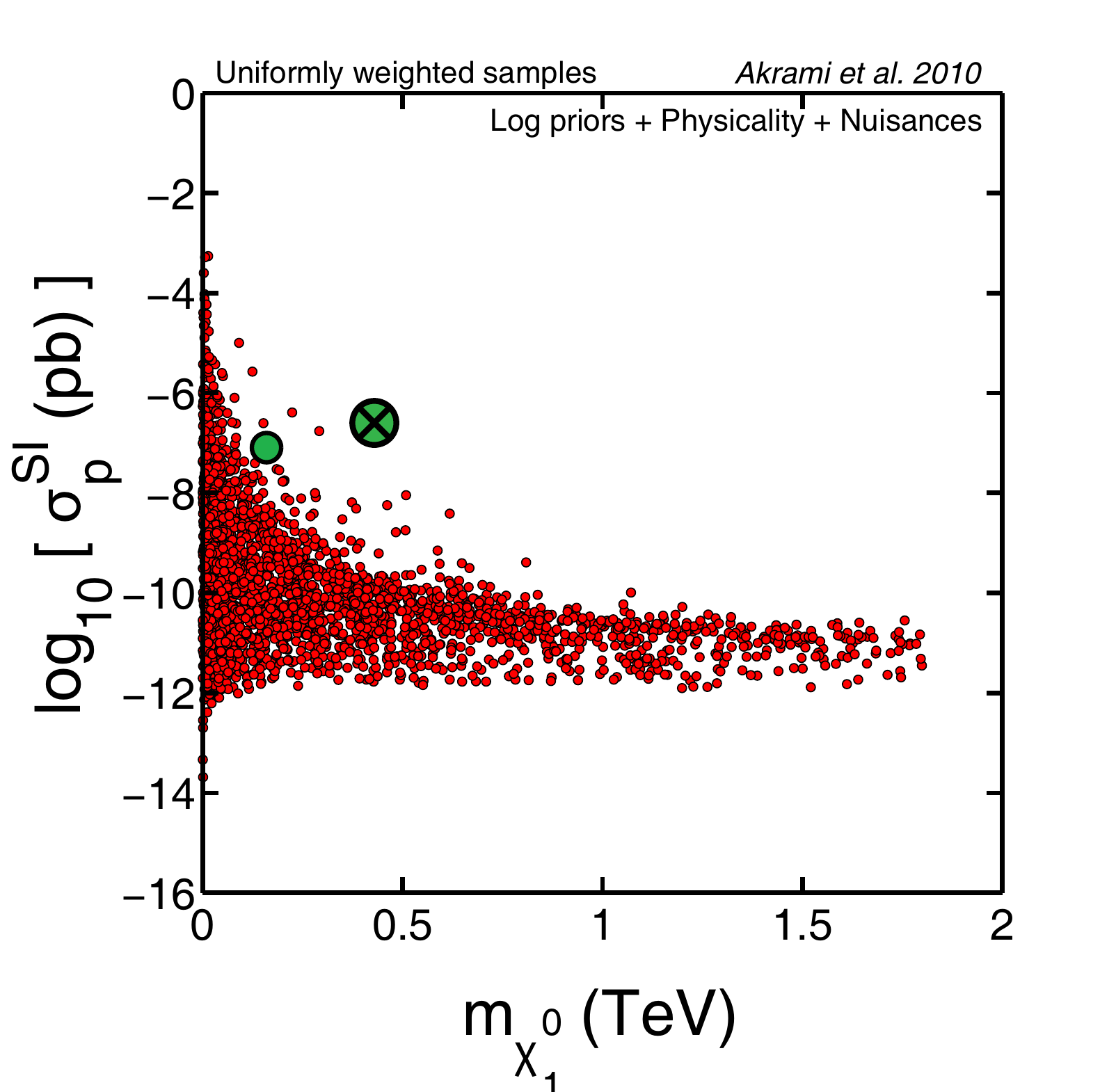}}\\
\caption[aa]{\footnotesize{Sample points for a scan with only physicality conditions and experimental constraints upon SM nuisance parameters, projected on the $m_{\tilde\chi^0_1}$-$\sigma^{SI}_p$ plane (DD constraints are not imposed here). Left and right panels represent scans with flat and log priors, respectively. Green dots and crosses show our benchmark points 1 and 2, respectively. Samples are plotted using a thinning factor of $10$.}}\label{fig:2D_phys_priors_DD}
\end{figure}

Fortunately, the particular likelihood function we have chosen for our analysis, depends mainly on two derived quantities, i.e. the neutralino mass $m_{\tilde\chi^0_1}$ and its spin-independent scattering cross-section with nucleons $\sigma^{SI}$. We therefore begin our discussion by showing 2D projections of the entire set of sample points on the $m_{\tilde\chi^0_1}$-$\sigma^{SI}_p$ plane, in~\fig{fig:2D_phys_priors_DD}. These samples are products of a scan with only physicality conditions and direct experimental constraints on the SM nuisance parameters. The left panel represents the distributions when flat priors are imposed on all parameters whereas the right panel does the same for the case with log priors imposed on $\mzero$ and $\mhalf$. Dots and crosses indicate locations of our benchmarks 1 and 2 on this plane, respectively.

In both flat and log-prior scans, ~\fig{fig:2D_phys_priors_DD} indicates that the effective prior on neutralino masses and SI scattering cross-sections is such that a large range of cross-sections is sampled at low neutralino masses.  In contrast, at high masses a much narrower range of cross-sections can be seen, centred on quite low values of $\sigma^{SI}_p$.

Such an effective prior mostly arises from the physicality requirement that the lightest neutralino be the LSP.  The LSP constraint means that in order to avoid a sfermion LSP, we require $m_{\tilde\chi^0_1} < m_{\tilde{f}}$ be satisfied for all sfermions.  For large neutralino masses, this excludes the existence of light squarks.  Because squark exchange in the $s$-channel is in general a major contributor to the amplitude of SI neutralino-quark scattering (see e.g. Ref.~\cite{DMJungman}), precluding small squark masses suppresses nuclear-scattering cross-sections.  Because no experimental constraints have been imposed on these scans, there are also basically no major contributions from special regions such as the focus point or Higgs funnel, where this deficit could potentially be compensated for by e.g. very low Higgs masses giving rise to large contributions to $\sigma^{SI}_p$ from Higgs-mediated diagrams.\footnote{Roughly speaking, large $m_{\tilde\chi^0_1}$ together with the neutralino LSP constraint also precludes low $m_0$, which tends to prevent very low Higgs masses, also suppressing the amplitude of Higgs exchange diagrams.} 

Because we also impose the EWSB physicality constraint, there is also an effective upper bound imposed on the mass of the lightest Higgs; this means that even for very large $m_{\tilde\chi^0_1}$, only relatively low Higgs masses are probed.  This means that even as $m_{\tilde\chi^0_1}$ continues to increase (implying an increase in the minimum values of $m_0$ permitted by the neutralino LSP constraint), the absolute contribution of Higgs exchange to $\sigma^{SI}_p$ stays roughly constant.  When combined with the suppression of the squark-exchange contribution, this means that $\sigma^{SI}_p$ becomes dominated by Higgs exchange at large $m_{\tilde\chi^0_1}$.  This produces a small band of permitted SI cross-sections at large $m_{\tilde\chi^0_1}$, which is approximately constant with increasing mass, and exhibits a width essentially set by the small allowed range of Higgs masses.

One feature of our results mentioned in the previous section is that the coverage for $m_{\tilde\chi^0_1}$ and $\sigma^{SI}_p$ is reduced by switching from benchmark 1 to benchmark 2, in both flat and log priors (\tab{tab:coverage}). This can be understood by comparing the locations of these two points in the $m_{\tilde\chi^0_1}$-$\sigma^{SI}_p$ planes of~\fig{fig:2D_phys_priors_DD}. The effective priors induced on these two quantities bias the samples towards lower values of $m_{\tilde\chi^0_1}$ and $\sigma^{SI}_p$. Our scanning algorithm tends to sample points around benchmark 1 better than around benchmark 2, as this bias is stronger further from the dense region of the plot. By having sampled the region around benchmark 1 more accurately than that around benchmark 2, we expect to find more high-likelihood points there with a direct impact on the constructed confidence regions, resulting in better coverage. This also means that such effects should be enhanced for log priors, because the concentration of points drops more sharply with increasing mass and cross-section in the log prior scan. This is seen in our results: coverage drops when changing from flat to log priors at each benchmark point individually, and also drops more severely at benchmark~2 than benchmark~1.

One potential reason for observing poor coverage in a statistical analysis of a complex model like the CMSSM is that the true parameters may lie at the boundary of the accessible parameter space (produced by imposing physically conditions) causing a poor convergence of the likelihood ratio test statistics (used to construct approximate confidence intervals) to its asymptotic chi-square distribution. In other words, Wilks' theorem does not apply in this case~\cite{wilks1,wilks2}. This effect is independent of the sampling accuracy and can happen even in the limit of good sampling. To see whether this is the case for our benchmarks i.e. whether they lie at the boundary or not, we have used a scan with artificial priors that force the scanning algorithm to sample regions of the parameter space in the vicinity of the benchmarks (i.e. the regions with rather high masses and cross-sections) with a higher resolution. We find a large cloud of CMSSM points in those regions with the benchmarks well inside the cloud. This clearly shows that the benchmarks do not lie at the boundary and there are many physical points around the benchmarks that have not been found by the normal usage of~\MN, i.e. when flat or log priors are used. In addition, the described boundary effects always lead to over-coverage (as is shown e.g. in Ref.~\cite{Bridges2010}). What we see is an opposite effect, present in parts of the parameter space where the neutralino LSP condition indirectly leads to under-sampling, as the (Bayesianally-optimised) scanning algorithm preferentially explores other parts of the parameter space. It therefore means that the under-coverage in our results is caused by poor sampling rather than the boundary effects.

\begin{table}[t]
\begin{center}
\begin{tabular}{|l | >{\centering\arraybackslash}p{1.2cm} | >{\centering\arraybackslash}p{1.6cm} | >{\centering\arraybackslash}p{1.6cm} | >{\centering\arraybackslash}p{1.6cm} | >{\centering\arraybackslash}p{1.6cm} |} \cline{3-6}
\multicolumn{2}{c|}{}& \multicolumn{4}{c|}{\textbf{Benchmark 2 (conf. int.)}} \bigstrut[t]\\ \cline{3-6}
\multicolumn{2}{c|}{}& \multicolumn{2}{c|}{\textbf{$\mu_S \simeq 13$}} & \multicolumn{2}{c|}{\textbf{$\mu_S \simeq 367$}} \bigstrut\\ \cline{3-6}
\multicolumn{2}{c|}{} & $1\sigma$ & $2\sigma$ & $1\sigma$ & $2\sigma$ \bigstrut\\ \hhline{--====}
\multirow{8}{*}{\begin{sideways}\textbf{Log priors}\end{sideways}} & $\mzero$          & \textcolor{red}{\textbf{17}} & \textcolor{red}{\textbf{47}}  &  \textcolor{green}{\textbf{75}} & \textcolor{green}{\textbf{100}} \bigstrut \\ \cline{2-6}\cline{2-6}
 & $\mhalf$          & \textcolor{red}{\textbf{1}} & \textcolor{red}{\textbf{17}}  &  \textcolor{red}{\textbf{35}} & \textcolor{red}{\textbf{80}} \bigstrut\\ \cline{2-6}
 & $\azero$          & \textcolor{black}{\textbf{91}} & \textcolor{black}{\textbf{100}}  &  \textcolor{black}{\textbf{100}} & \textcolor{green}{\textbf{100}} \bigstrut\\ \cline{2-6}
 & $\tanb$           & \textcolor{black}{\textbf{99}} & \textcolor{black}{\textbf{100}}&  \textcolor{black}{\textbf{100}} & \textcolor{green}{\textbf{100}} \bigstrut\\ \cline{2-6}
 & $m_{\tilde\chi^0_1}$          & \textcolor{red}{\textbf{2}} & \textcolor{red}{\textbf{15}} &  \textcolor{red}{\textbf{45}} & \textcolor{red}{\textbf{85}} \bigstrut\\\cline{2-6}
 & $\sigma^{SI}_p$          & \textcolor{red}{\textbf{15}} & \textcolor{red}{\textbf{59}} &  \textcolor{black}{\textbf{85}} & \textcolor{green}{\textbf{100}} \bigstrut\\ 
\hline
\end{tabular}
\caption[aa]{\footnotesize{Comparison of our coverage results with low and high statistics. Here, the true parameters are set to those of benchmark 2, log priors are imposed on $\mzero$ and $\mhalf$ and the coverage is evaluated for $1\sigma$ and $2\sigma$ confidence intervals. The low-statistics case (first two data columns from the left) is the same as the corresponding original coverage study in~\tab{tab:coverage}, and the high-statistics case (last two data columns from the left) is obtained by increasing the XENON exposure by a factor of 30. The quantity $\mu_S$ in each case is the expected number of signal events, which is about 13 and 367 for low and high statistics, respectively. As in~\tab{tab:coverage}, numbers in red and black show under- and over-coverage, respectively, while the ones in green show correct coverage. The high-statistics results are based on 20 trials and the binomial estimate of errors implies that the green numbers in this case are the ones that fall within $[58,78]$ and $[91,100]$ intervals for $1\sigma$ and $2\sigma$ confidence levels, respectively.}} \label{tab:higherstats}
\end{center}
\end{table}

Additionally, because the sampling effect is the dominant effect in our results, we expect that higher statistics lead to better coverage, as they lead to a reduced influence of the effective priors placed on the scanning algorithm by the boundary conditions. We have tested this by performing a coverage study for benchmark 2 using the same XENON10-like likelihood but with 30 times larger exposure. This provides substantially larger statistics (for benchmark 2, for instance, the expected number of signal events ($\mu_S$) becomes about 367 as opposed to about 13 for the lower-statistics case). We use log priors for these scans and compare our results with the corresponding ones in \tab{tab:coverage} (benchmark 2, confidence intervals and log priors). Due to the computational limitations, we have only performed 20 scans in this case and evaluated the coverage based on them. This means that in this case using the binomial estimate of errors implies that a count of successes for a $1\sigma$ ($2\sigma$) confidence level shows exact coverage if it lies in the range $[58,78]$ ($[91,100]$). The new degrees of coverage compared to the current ones are given in~\tab{tab:higherstats}. The significant improvement in the coverage for $\mzero$, $\mhalf$, $m_{\tilde\chi^0_1}$ and $\sigma^{SI}_p$ clearly shows that for the same set of true parameters and by only increasing the statistics, one gets substantially better coverage in these cases.

It is important to realise the consequences of~\fig{fig:2D_phys_priors_DD} and our above interpretation of coverage in terms of sampling effects in the context of future attempts to constrain SUSY parameters with large-scale DD experiments~\cite{Akrami:2010}. Since effective priors force the scanning algorithm to explore lower masses and cross-sections with a larger resolution, one expects much more accurate likelihood mappings for those regions. The coverage is thus less of an issue for smaller cross-sections. This is interesting because if WIMPs exist and they happen to have very low nuclear scattering cross-sections, even future ton-scale experiments will be able to detect only a few recoil events. This means that the statistics will be very low and, consequently, without having sufficiently high-resolution likelihood mappings of regions close to the actual point, any attempts at making proper statistical inference will be in vain.

\begin{table}[t]
\begin{center}
{\small
\begin{tabular}{p{3.5cm} r}
\toprule
\textbf{Quantity}& \multicolumn{1}{c}{\textbf{Benchmark 3}} \\ \toprule
$\mzero$ (GeV)           & $3834.35$ \\
$\mhalf$ (GeV)       & $378.02$ \\
$\azero$ (GeV)          & $2124.04$ \\
$\tanb$     & $22.46$ \\ \midrule
$\mtpole$ (GeV)      & $174.38$ \\
$m_b (m_b)^{\overline{MS}}$ (GeV) & $4.10$ \\
$\alphas$       & $0.1140$ \\
$1/\alphaemmz$  & $127.926$ \\ \midrule
$\sigma^{SI}_p$ (pb) & 1.0 $\times 10^{-10}$ \\
$m_{\tilde\chi^0_1}$ (GeV)  & $159.87$ \\ \bottomrule
\end{tabular}
} \caption[aa]{\footnotesize{Values of model and nuisance parameters at benchmark 3, as well as the corresponding DD quantities $\sigma^{SI}_p$ and $m_{\tilde\chi^0_1}$.}} \label{tab:params4BM3}
\end{center}
\end{table}

We have tested this conjecture by performing another coverage study for a benchmark with $m_{\tilde\chi^0_1}$ and $\sigma^{SI}_p$ lying within the high-sample region in~\fig{fig:2D_phys_priors_DD}, i.e. with a low neutralino mass and a small cross-section. Values of the CMSSM parameters, as well as the nuisance parameters and the DD quantities $\sigma^{SI}_p$ and $m_{\tilde\chi^0_1}$ at this benchmark 3 are given in~\tab{tab:params4BM3}. We present our coverage results in~\tab{tab:benchmark3}, where they are provided for confidence intervals and for cases of low and high statistics. Log priors have been used in the scans and the results are based on 100 trials for each case. The first two data columns, correspond to the low-statistics analysis in which our original XENON likelihood with an exposure of 1000 kg-days is used. The number of expected signal events $\mu_S$ in this case becomes effectively zero. For the high statistics case (last two data columns), we have increased the XENON exposure by a factor of $900$ so as to get approximately the same number of expected signal events ($\sim 13$) as in our previous cases of benchmarks 1 and 2. This makes the comparison of the cases easier. Our results show over-coverage for benchmark 3 (for both low and high statistics) and the coverage appears to be substantially higher (for most quantities) than the cases for benchmarks 1 and 2. The overall significant over-coverage observed here confirms our interpretations about the roles of sampling deficiency in obtaining poor coverage. This is because, as we see here, by moving to the high-sample region of~\fig{fig:2D_phys_priors_DD}, the high concentration of sample points at low cross-sections (and low masses) provides the high-resolution mapping required for having a good coverage. The reason for observing higher levels of over-coverage in the low-statistics case of~\tab{tab:benchmark3} compared to the high-statistics one, is also understood. The absence of any signal events in most realisations of the experiment in the former case (with $\mu_S=0$) means that the constraints placed on the parameters by the data are so weak that the constructed confidence intervals become very large and almost always contain the true parameters. This clearly leads to a significant over-coverage.

\begin{table}[t]
\begin{center}
\begin{tabular}{|l | >{\centering\arraybackslash}p{1.2cm} | >{\centering\arraybackslash}p{1.6cm} | >{\centering\arraybackslash}p{1.6cm} | >{\centering\arraybackslash}p{1.6cm} | >{\centering\arraybackslash}p{1.6cm} |} \cline{3-6}
\multicolumn{2}{c|}{}& \multicolumn{4}{c|}{\textbf{Benchmark 3 (conf. int.)}} \bigstrut[t]\\ \cline{3-6}
\multicolumn{2}{c|}{}& \multicolumn{2}{c|}{\textbf{$\mu_S \simeq 0$}} & \multicolumn{2}{c|}{\textbf{$\mu_S \simeq 13$}} \bigstrut\\ \cline{3-6}
\multicolumn{2}{c|}{} & $1\sigma$ & $2\sigma$ & $1\sigma$ & $2\sigma$ \bigstrut\\ \hhline{--====}
\multirow{8}{*}{\begin{sideways}\textbf{Log priors}\end{sideways}} & $\mzero$           &  \textcolor{black}{\textbf{99}} & \textcolor{black}{\textbf{100}} & \textcolor{black}{\textbf{100}} & \textcolor{black}{\textbf{100}} \bigstrut \\ \cline{2-6}\cline{2-6}
 & $\mhalf$           &  \textcolor{black}{\textbf{99}} & \textcolor{black}{\textbf{100}} & \textcolor{black}{\textbf{73}} & \textcolor{black}{\textbf{99}} \bigstrut\\ \cline{2-6}
 & $\azero$           &  \textcolor{black}{\textbf{99}} & \textcolor{black}{\textbf{100}} & \textcolor{black}{\textbf{89}} & \textcolor{black}{\textbf{100}} \bigstrut\\ \cline{2-6}
 & $\tanb$           &  \textcolor{black}{\textbf{99}} & \textcolor{black}{\textbf{100}} & \textcolor{black}{\textbf{99}} & \textcolor{black}{\textbf{100}} \bigstrut\\ \cline{2-6}
 & $m_{\tilde\chi^0_1}$           &  \textcolor{black}{\textbf{99}} & \textcolor{black}{\textbf{100}} & \textcolor{black}{\textbf{76}} & \textcolor{black}{\textbf{98}} \bigstrut\\\cline{2-6}
 & $\sigma^{SI}_p$           &  \textcolor{black}{\textbf{98}} & \textcolor{black}{\textbf{98}} & \textcolor{black}{\textbf{76}} & \textcolor{black}{\textbf{99}} \bigstrut\\ 
\hline
\end{tabular}
\caption[aa]{\footnotesize{Results of our coverage evaluations for benchmark 3 (\tab{tab:params4BM3}) with a low neutralino mass and a small cross-section located within the high-sample region of~\fig{fig:2D_phys_priors_DD}. Here, log priors are imposed on $\mzero$ and $\mhalf$ and the coverage is evaluated for $1\sigma$ and $2\sigma$ confidence intervals. The results are given for both low and high statistics. The low-statistics case (first two data columns from the left) corresponds to the same XENON likelihood as used in~\tab{tab:coverage}, and the high-statistics case (last two data columns from the left) is obtained by increasing the XENON exposure by a factor of 900. The quantity $\mu_S$ in each case is the expected number of signal events, which is about 0 and 13 for low and high statistics, respectively. All numbers in this case show over-coverage.}} \label{tab:benchmark3}
\end{center}
\end{table}

Fortunately for WIMPs with high cross-sections, on the other hand, the large number of signal events at future detectors provides sufficiently high statistics to overcome sampling effects (as we discussed earlier).  Prospects for future SUSY parameter estimation at DD experiments would thus appear quite rosy.

In comparison to the recent complementary CMSSM coverage study~\cite{Bridges2010}, based on reconstruction of an ATLAS benchmark point with mock ATLAS results, we find rather different coverage properties.  Here, we find that intervals undercover for both our benchmark points 1 and 2, whilst Ref.~\cite{Bridges2010} find overcoverage for their chosen benchmark and pseudo-data.  Regarding our above discussion, the difference can likely be explained by the different locations of the chosen benchmarks; their benchmark lies deep in the well-sampled, low-mass part of the mSUGRA/CMSSM ``bulk'' region. As we argued above, the high concentration of samples in these regions greatly reduces the influence of sampling effects.  This in turn allows boundary effects to dominate. The over-coverage observed in these cases is then, as also argued by the authors of Ref.~\cite{Bridges2010}, likely to originate from the physical boundaries in the parameter space.

\begin{figure}[t]
\centering
\subfigure[][]{\includegraphics[width=0.34\linewidth, trim = 45 258 30 155, clip=true]{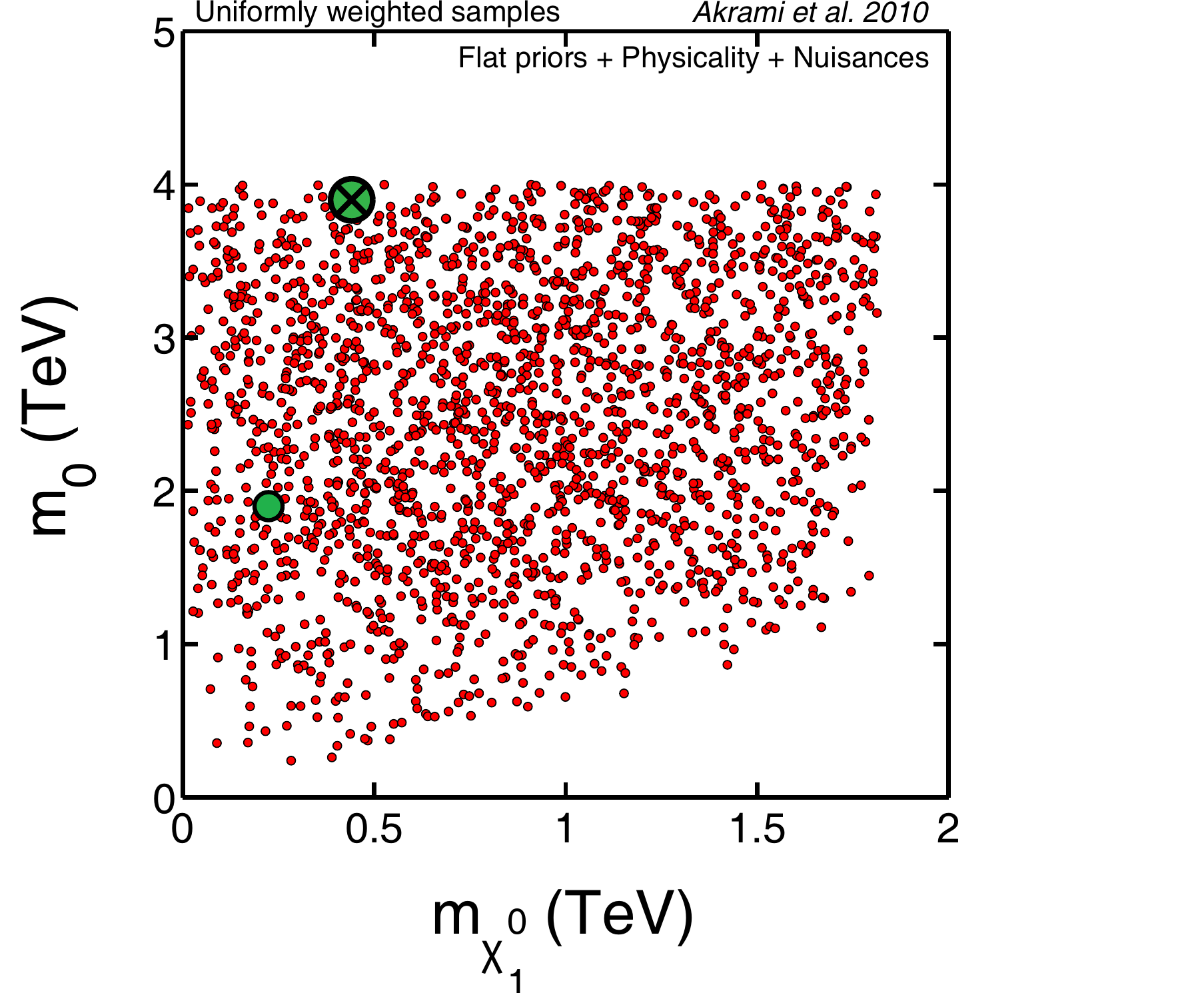}}
\subfigure[][]{\includegraphics[width=0.34\linewidth, trim = 45 258 30 155, clip=true]{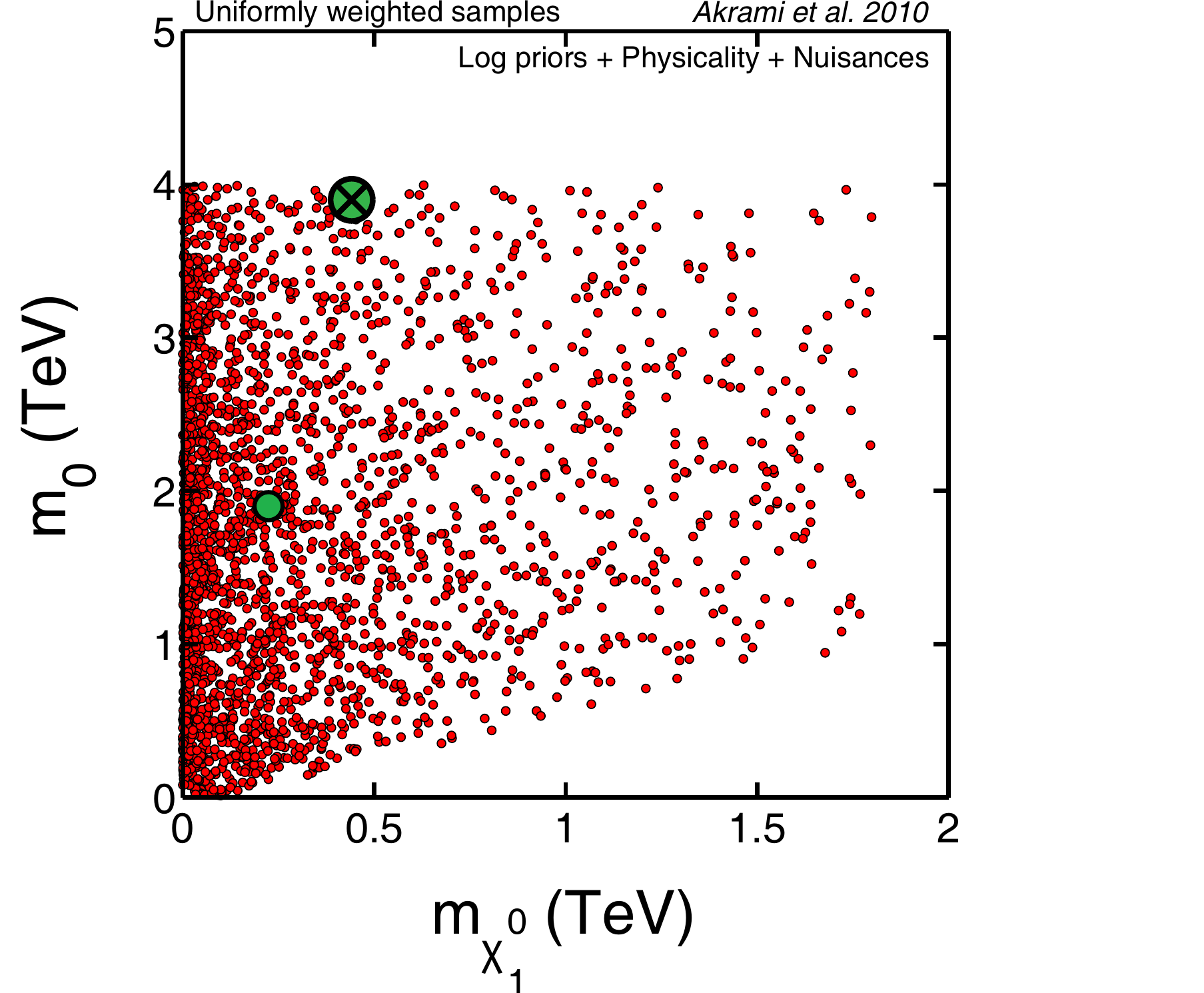}}\\
\subfigure[][]{\includegraphics[width=0.34\linewidth, trim = 45 258 30 155, clip=true]{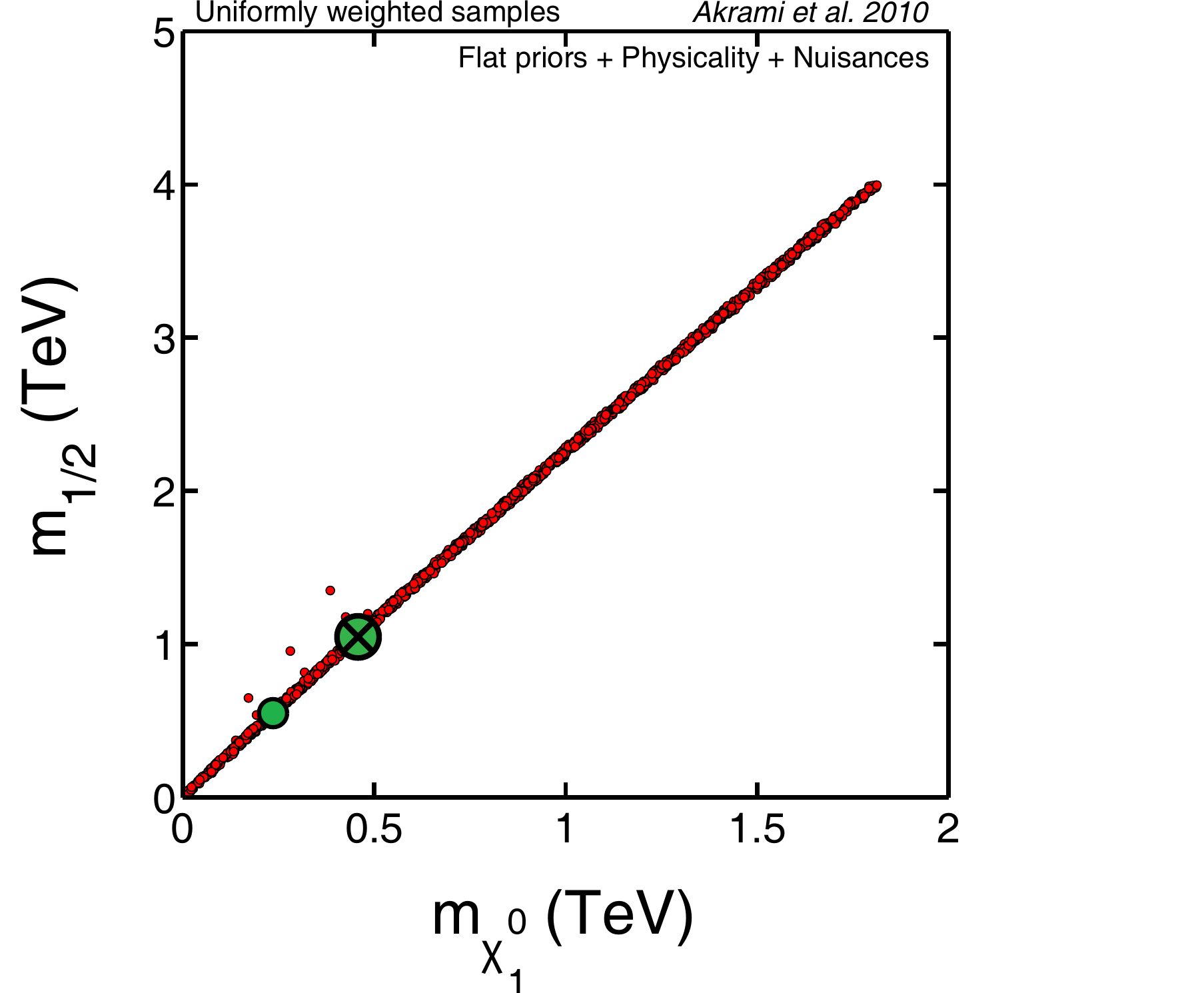}}
\subfigure[][]{\includegraphics[width=0.34\linewidth, trim = 45 258 30 155, clip=true]{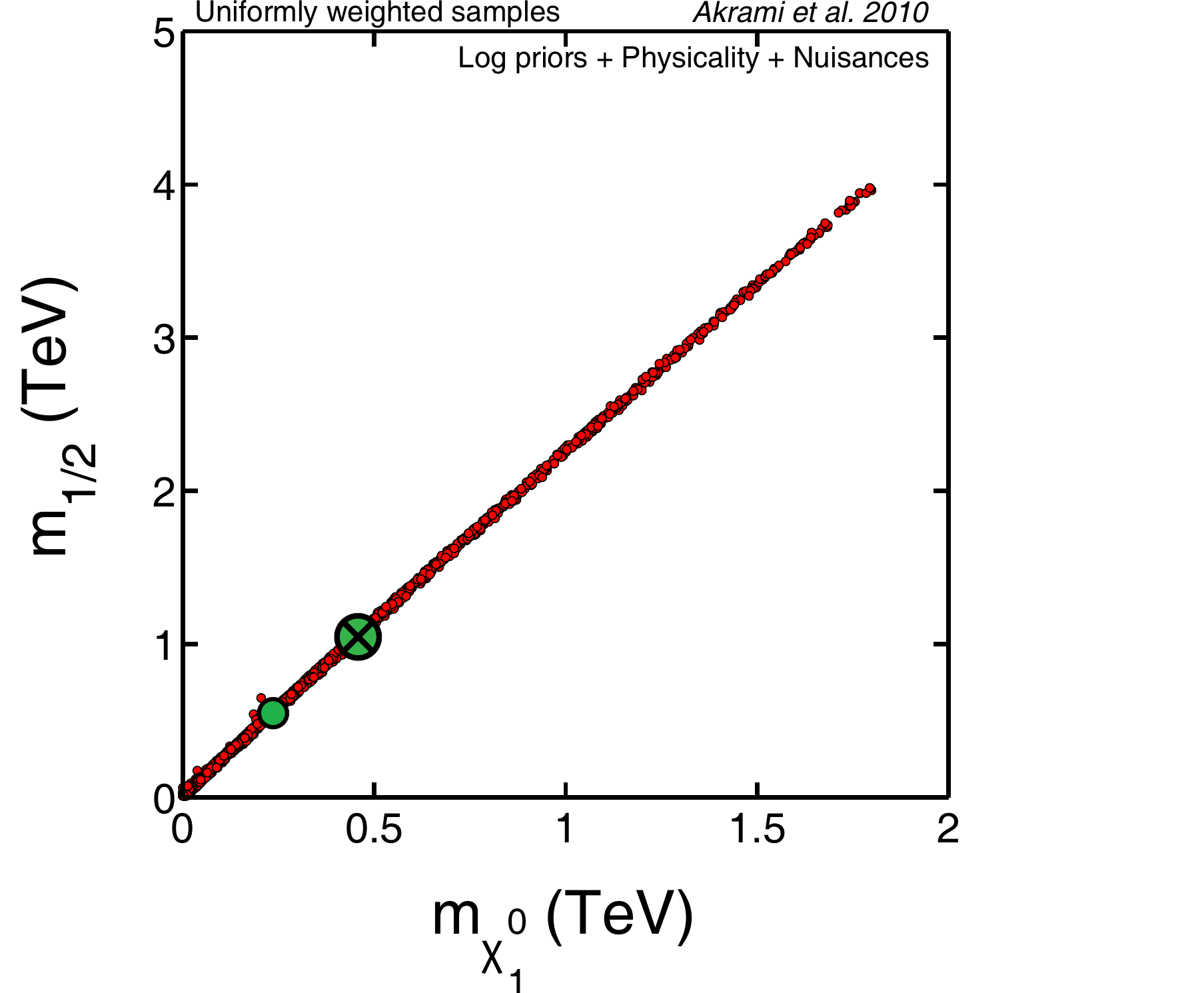}}\\
\subfigure[][]{\includegraphics[width=0.34\linewidth, trim = 45 258 30 155, clip=true]{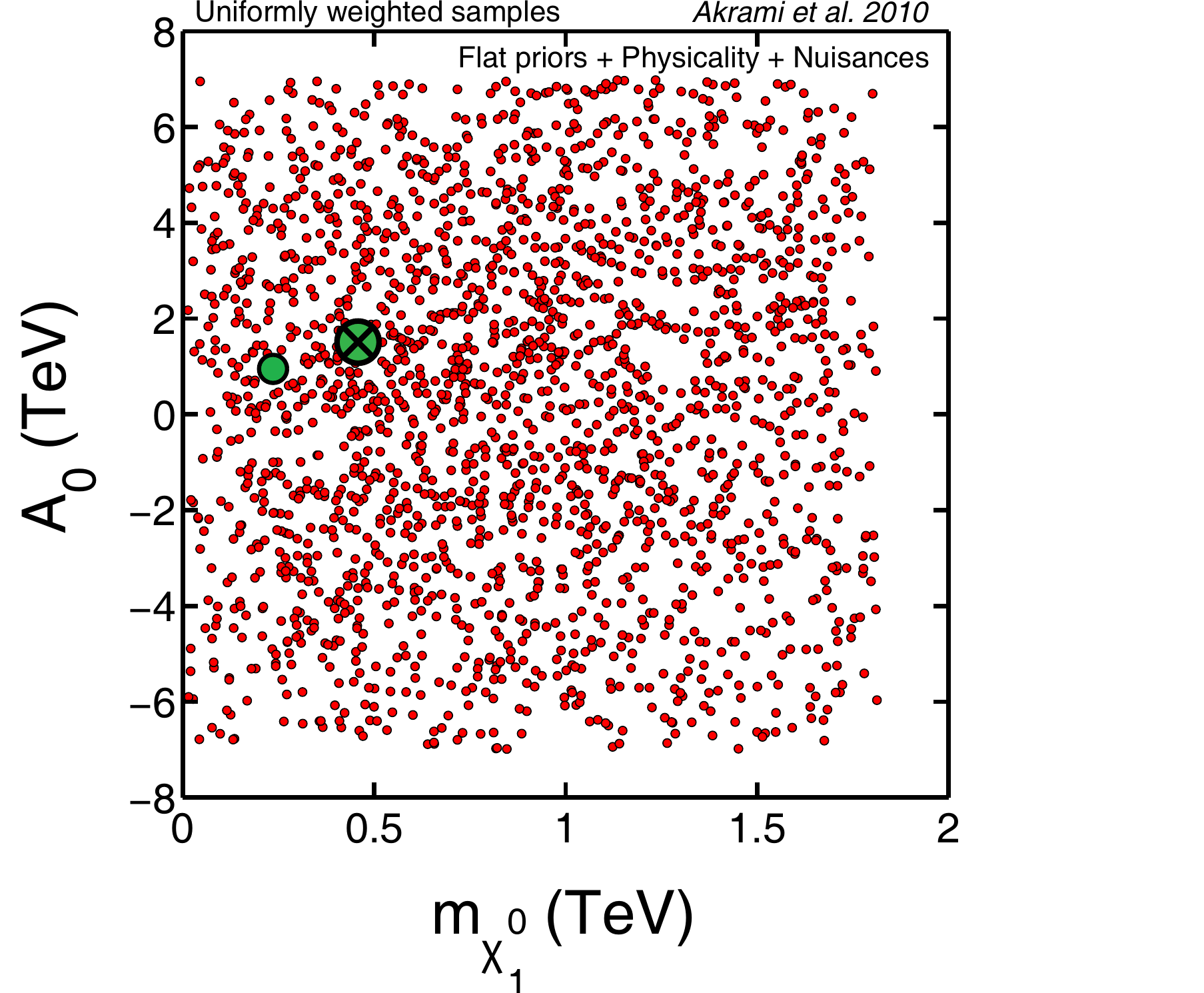}}
\subfigure[][]{\includegraphics[width=0.34\linewidth, trim = 45 258 30 155, clip=true]{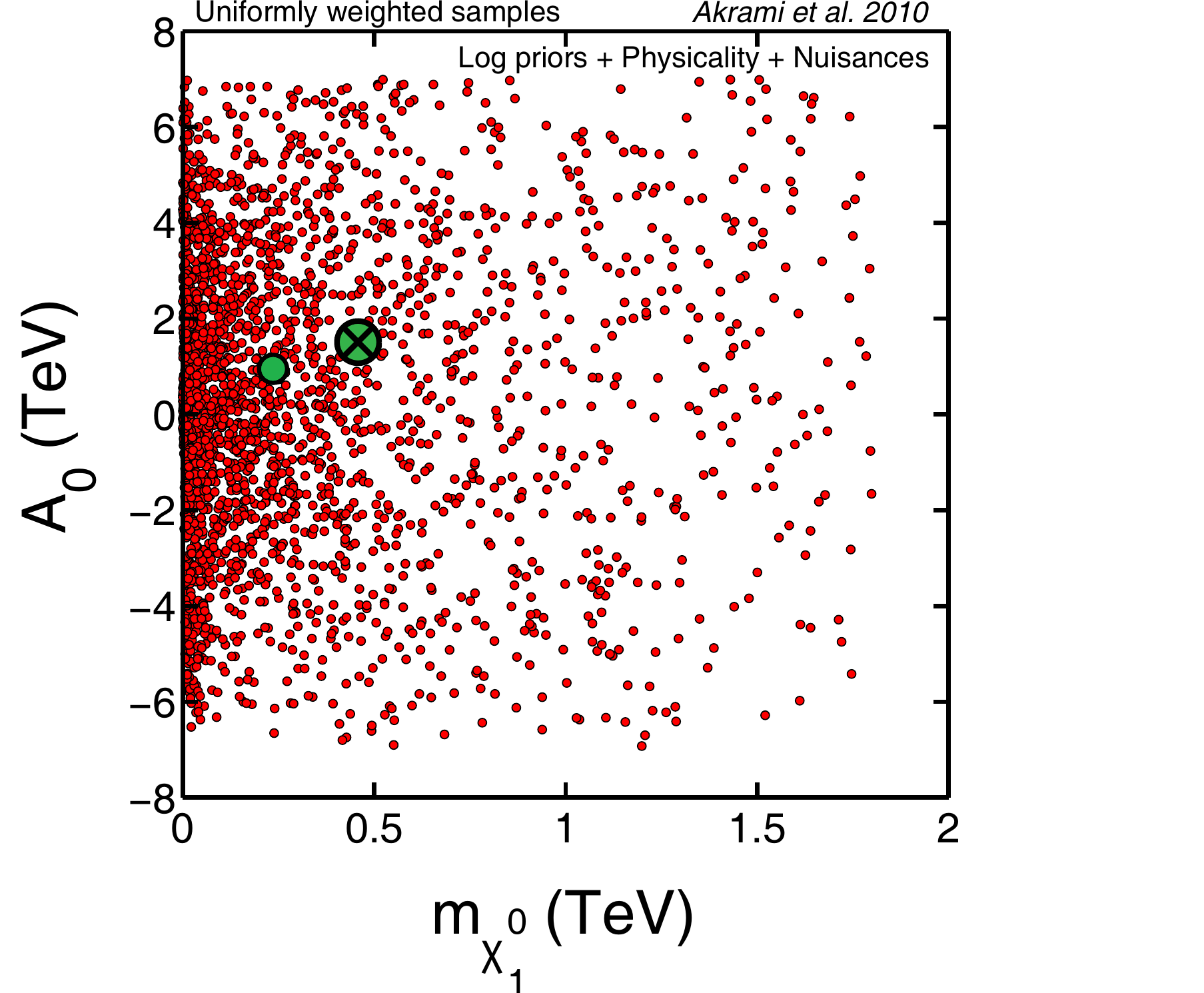}}\\
\subfigure[][]{\includegraphics[width=0.34\linewidth, trim = 45 258 30 155, clip=true]{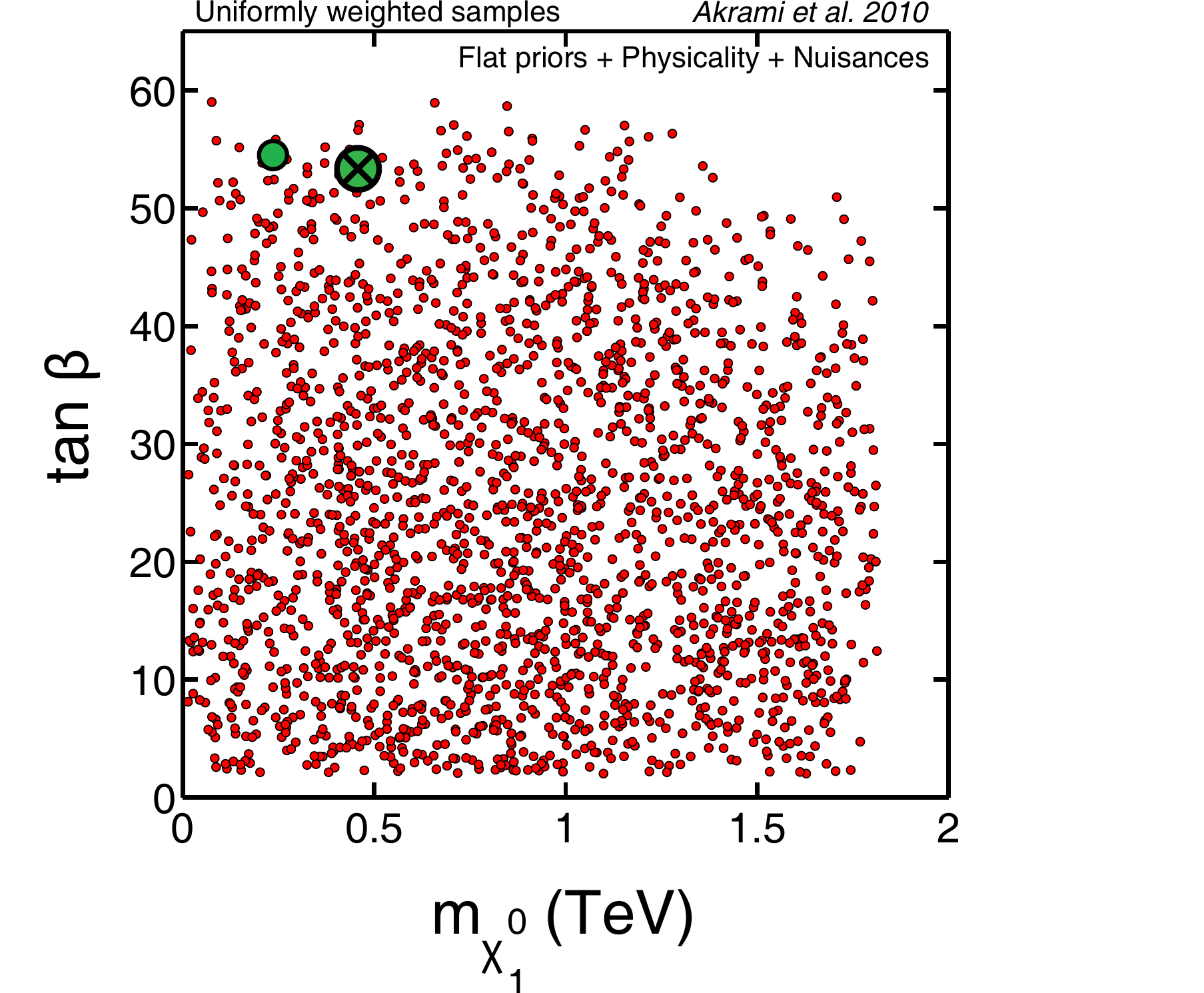}}
\subfigure[][]{\includegraphics[width=0.34\linewidth, trim = 45 258 30 155, clip=true]{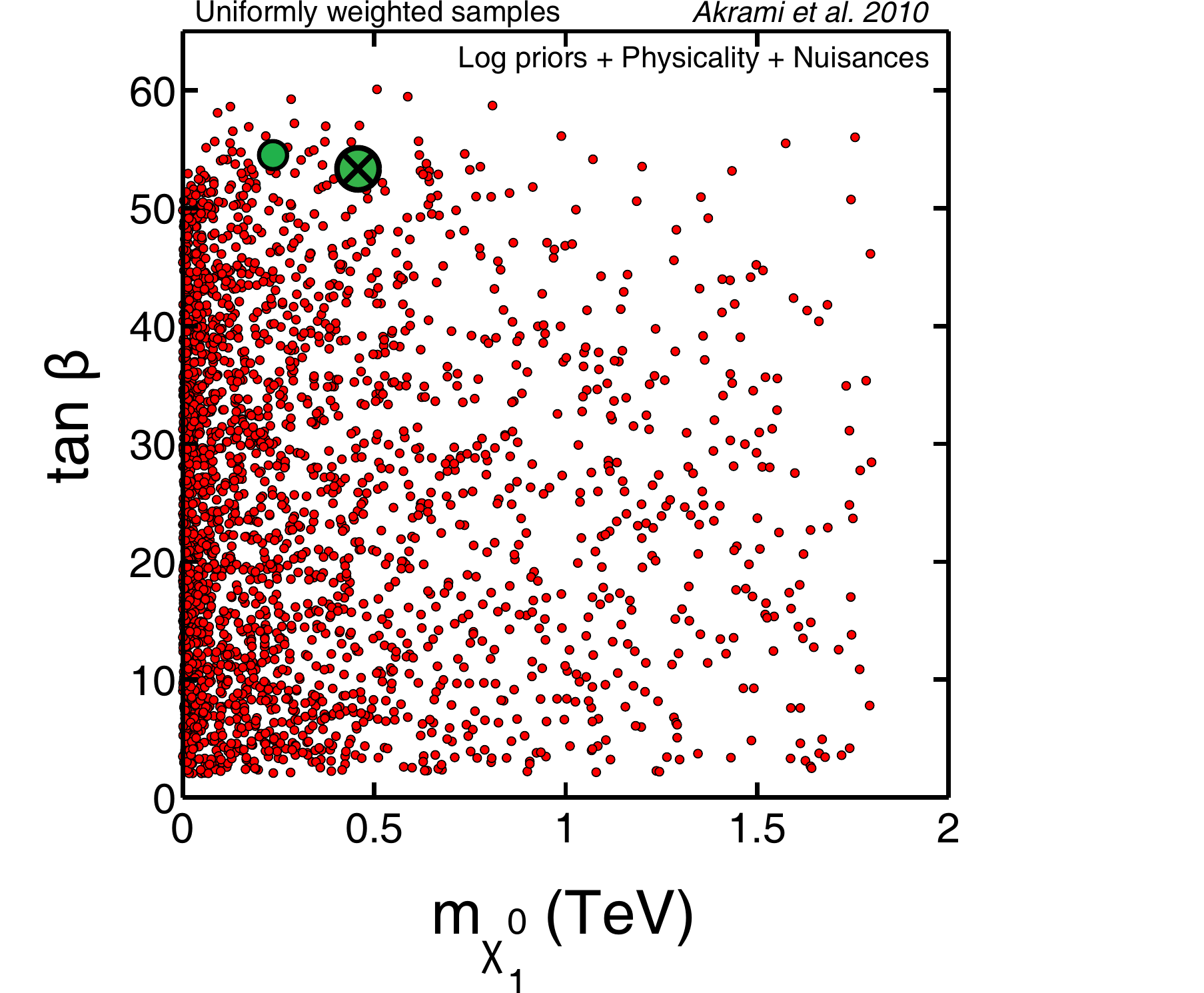}}\\
\caption[aa]{\footnotesize{As in~\fig{fig:2D_phys_priors_DD}, but for the CMSSM parameters $\mzero$, $\mhalf$, $\azero$ and $\tanb$ versus the neutralino mass $m_{\tilde\chi^0_1}$.}}\label{fig:2D_phys_priors_CMSSMmChi}
\end{figure}

\begin{figure}[t]
\centering
\subfigure[][]{\includegraphics[width=0.34\linewidth, trim = 45 258 30 155, clip=true]{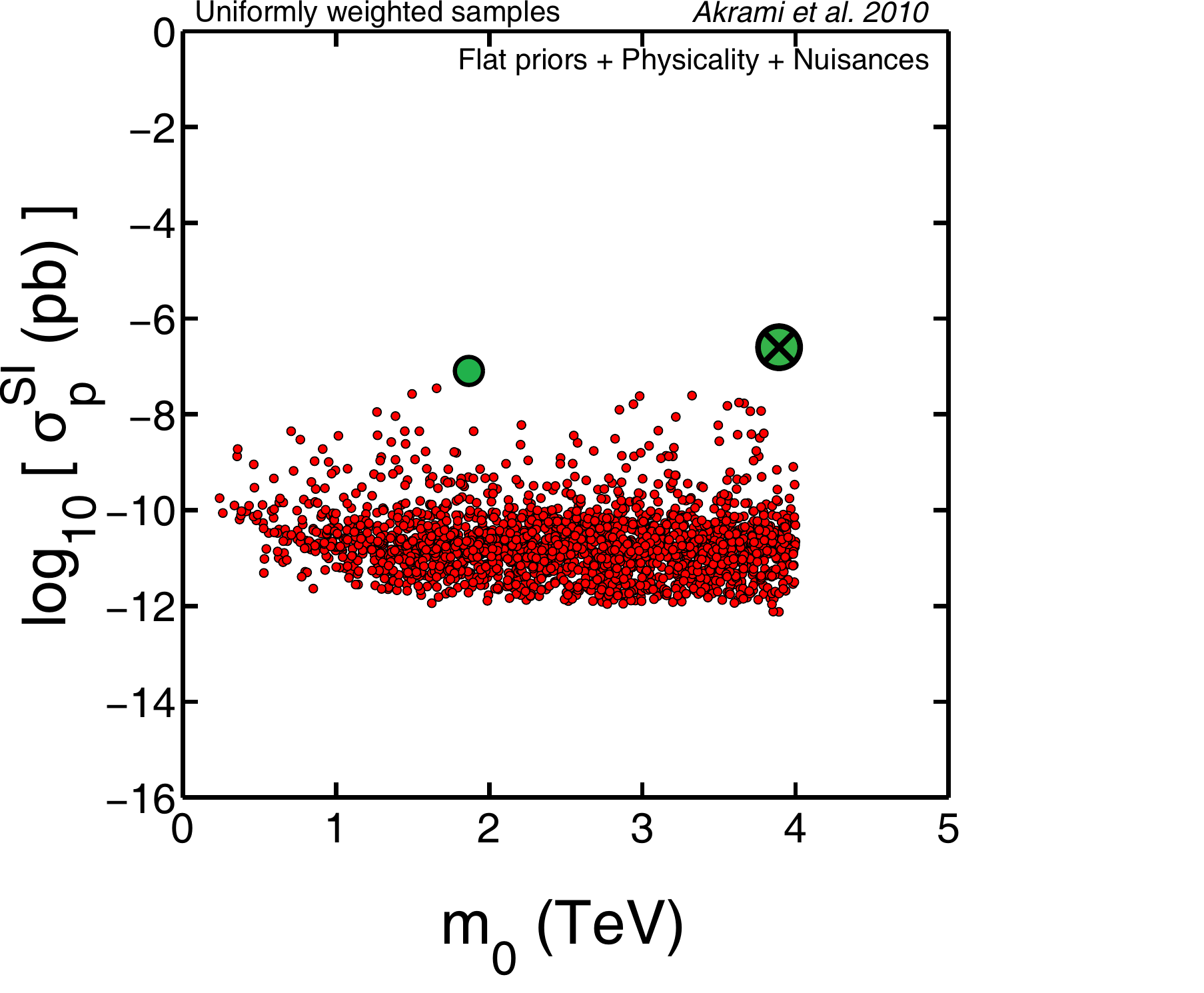}}
\subfigure[][]{\includegraphics[width=0.34\linewidth, trim = 45 258 30 155, clip=true]{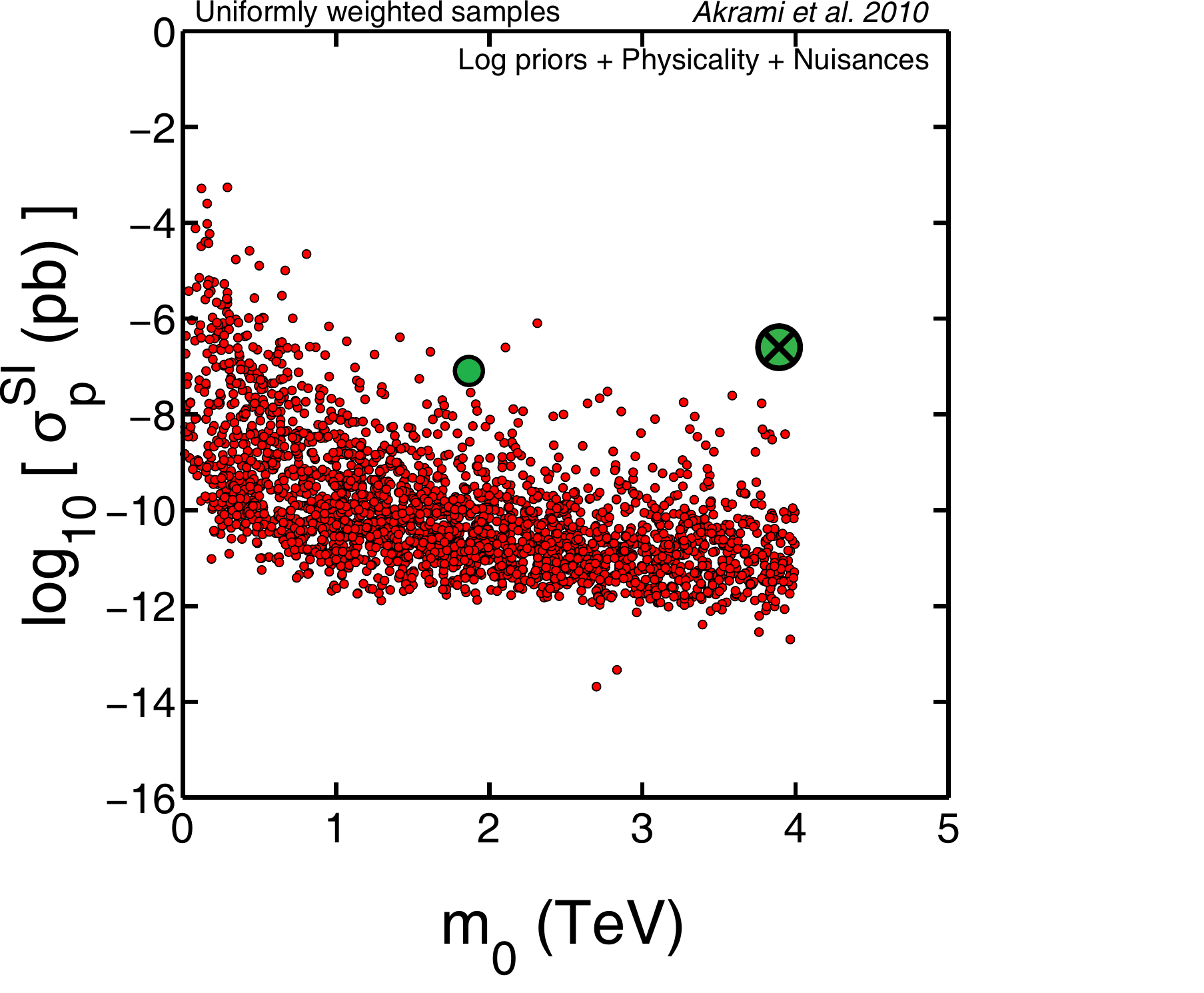}}\\
\subfigure[][]{\includegraphics[width=0.34\linewidth, trim = 45 258 30 155, clip=true]{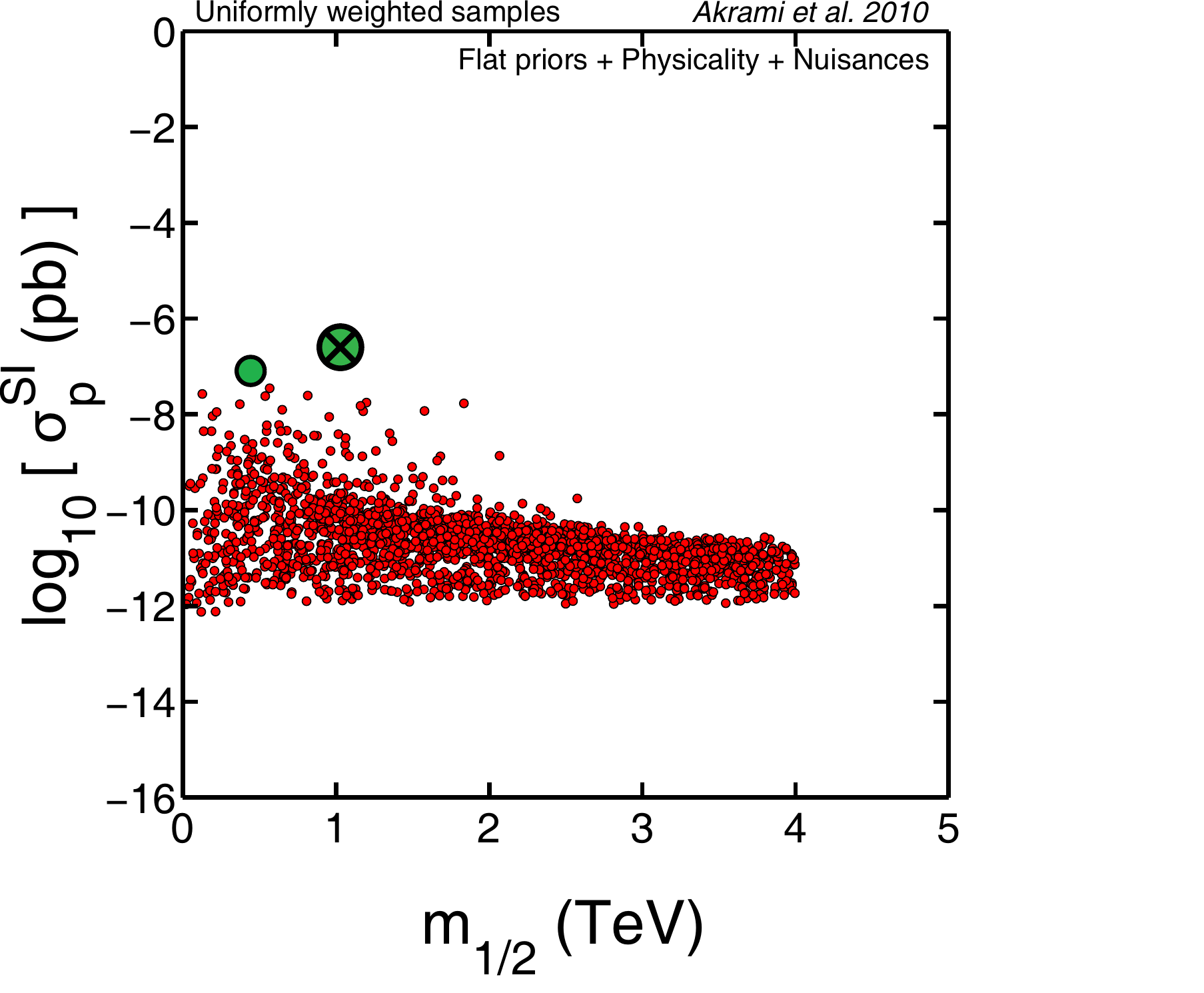}}
\subfigure[][]{\includegraphics[width=0.34\linewidth, trim = 45 258 30 155, clip=true]{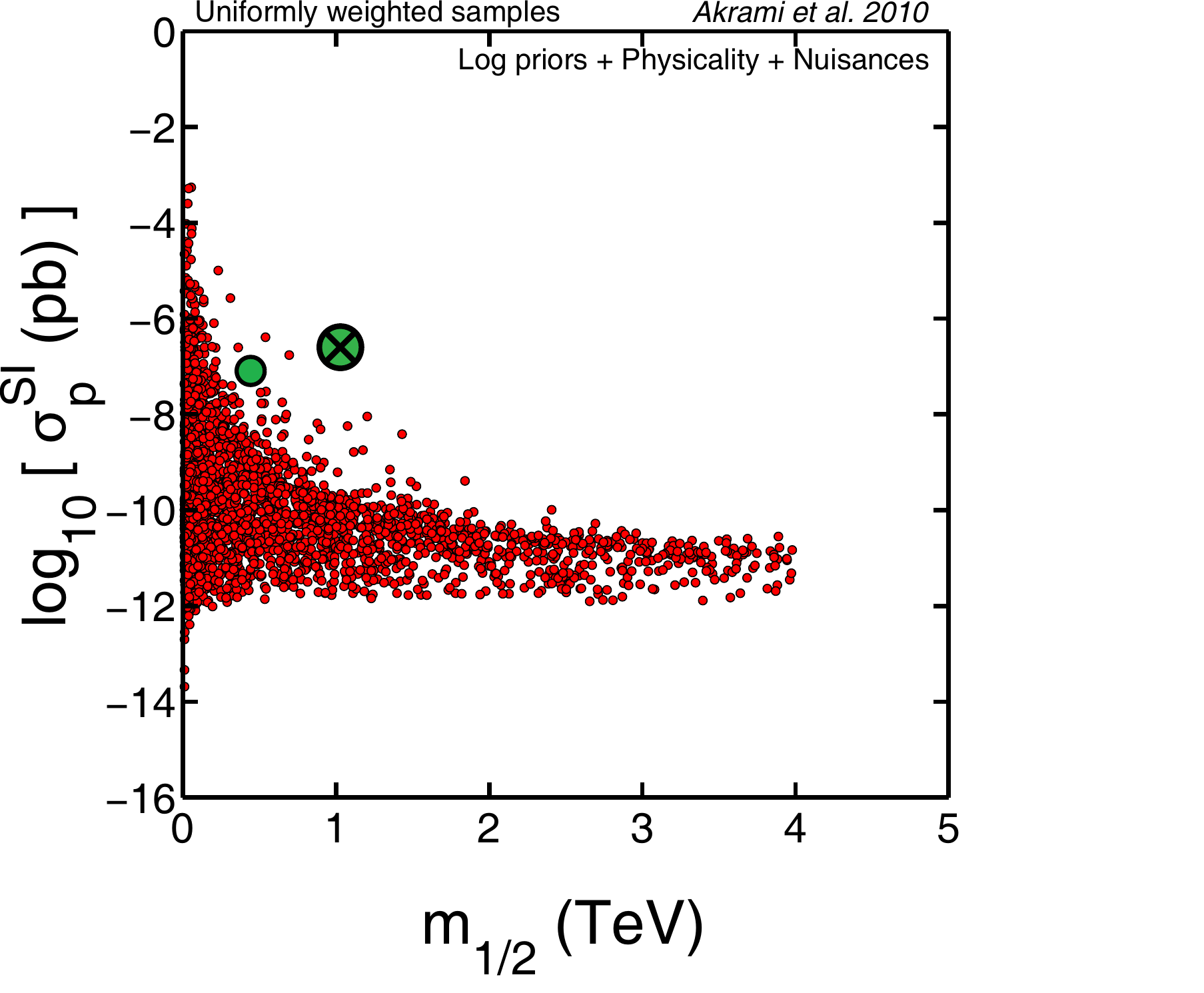}}\\
\subfigure[][]{\includegraphics[width=0.34\linewidth, trim = 45 258 30 155, clip=true]{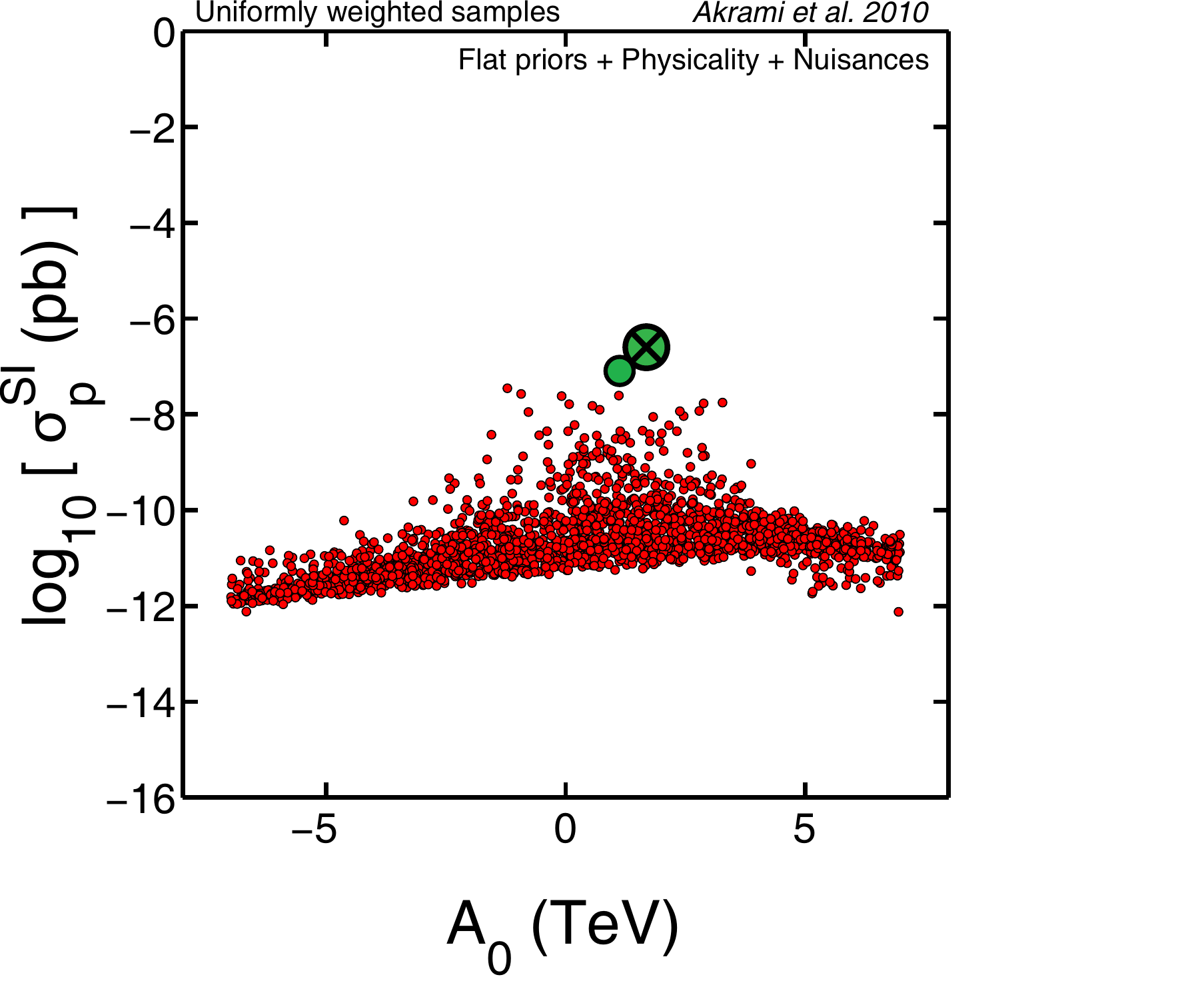}}
\subfigure[][]{\includegraphics[width=0.34\linewidth, trim = 45 258 30 155, clip=true]{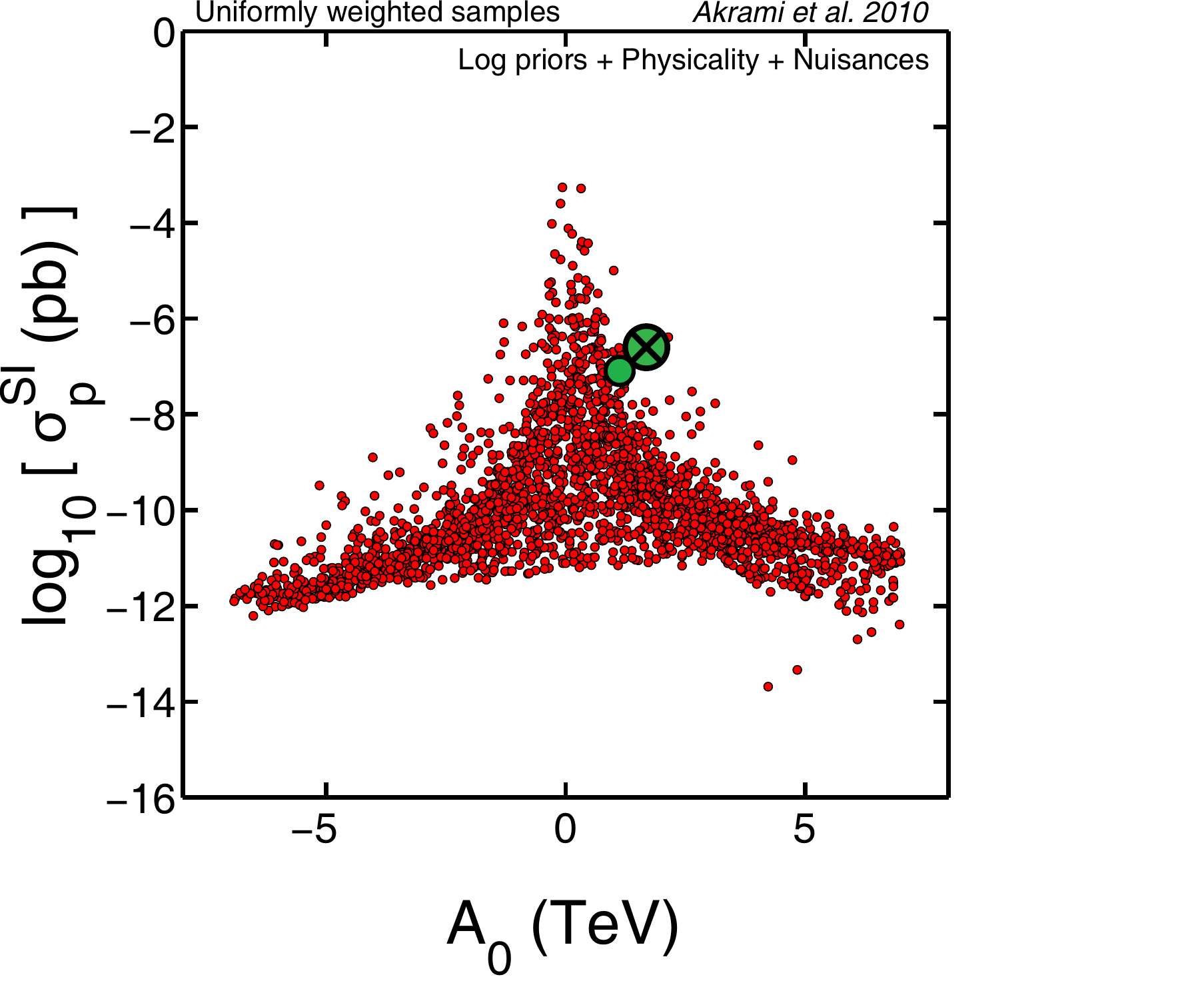}}\\
\subfigure[][]{\includegraphics[width=0.34\linewidth, trim = 45 258 30 155, clip=true]{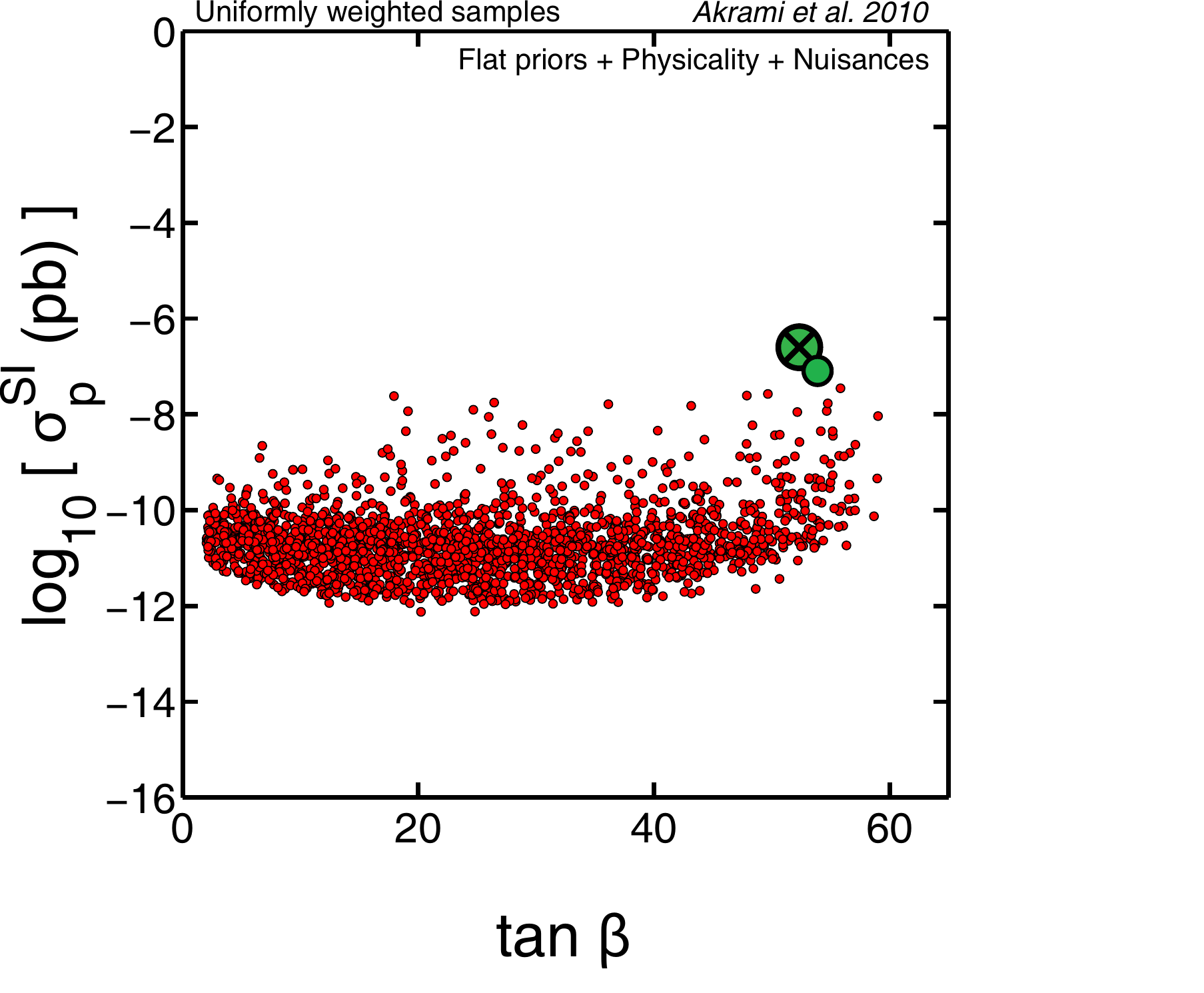}}
\subfigure[][]{\includegraphics[width=0.34\linewidth, trim = 45 258 30 155, clip=true]{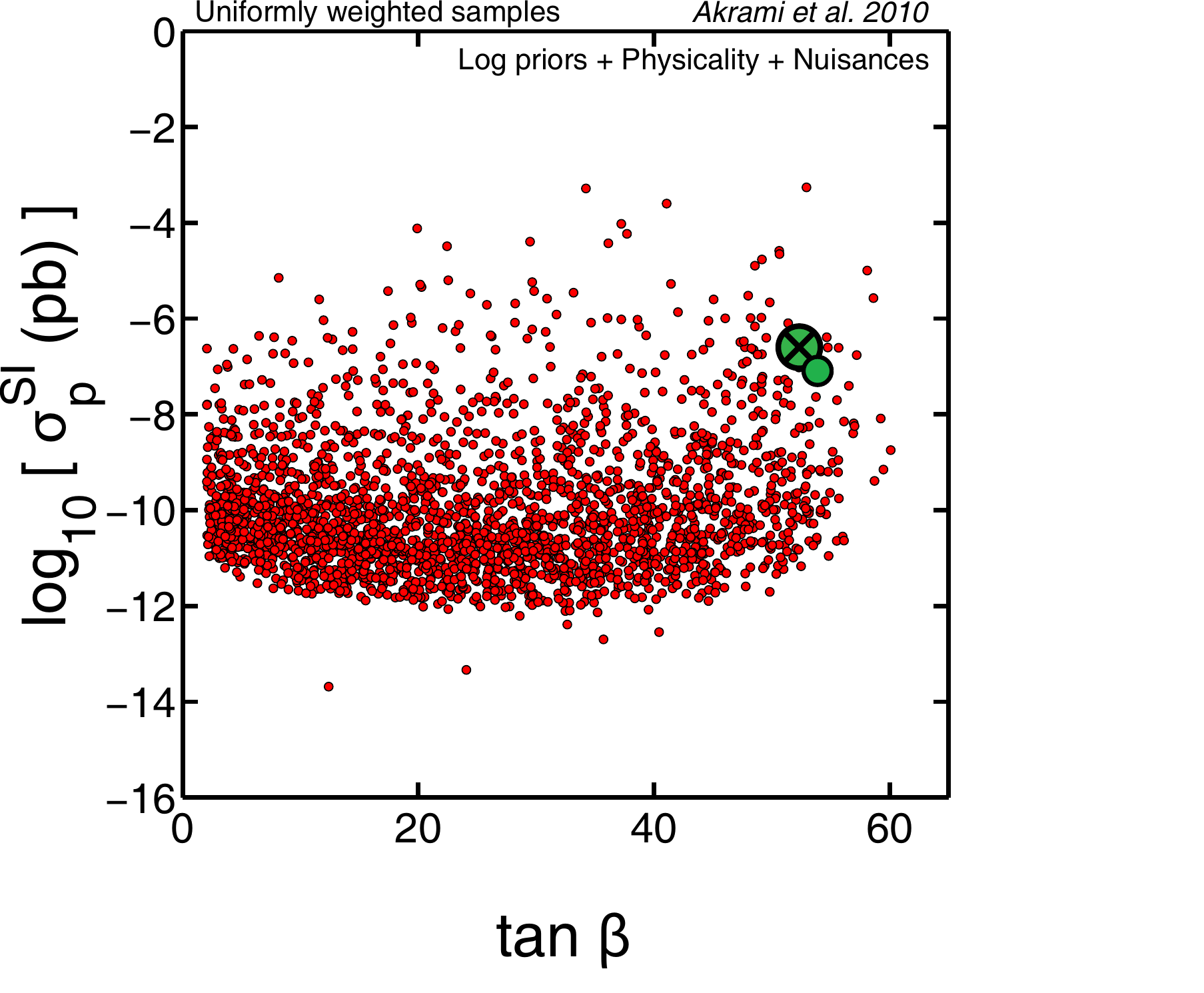}}\\
\caption[aa]{\footnotesize{As in~\fig{fig:2D_phys_priors_DD} and~\fig{fig:2D_phys_priors_CMSSMmChi}, but for the CMSSM parameters $\mzero$, $\mhalf$, $\azero$ and $\tanb$ versus the spin-independent scattering cross-section of the neutralino and a proton $\sigma^{SI}_p$.}}\label{fig:2D_phys_priors_CMSSMsigmaSI}
\end{figure}

In addition to the DD quantities $m_{\tilde\chi^0_1}$ and $\sigma^{SI}_p$, we see variations in coverage also for the CMSSM parameters, $\mzero$, $\mhalf$, $\azero$ and $\tanb$, when we change our benchmark or priors (\tab{tab:coverage}). Most of the patterns we observe here can be explained by analysing 2D plots of the parameters versus the observables $m_{\tilde\chi^0_1}$ and $\sigma^{SI}_p$. The sampling effects we discussed in terms of the DD quantities can propagate throughout the entire parameter space, impacting statistical inferences for other quantities like the CMSSM parameters themselves. To demonstrate these effects, let us therefore look at~\figs{fig:2D_phys_priors_CMSSMmChi}{fig:2D_phys_priors_CMSSMsigmaSI}, where samples for the scan with only priors, physicality and nuisances are given in planes of parameters versus $m_{\tilde\chi^0_1}$ and $\sigma^{SI}_p$. Again, left and right panels correspond to flat and log priors, respectively, and our two benchmarks (1 and 2) are shown by dots and crosses.

The gaugino mass parameter $\mhalf$ is perhaps the most obvious case.~\tab{tab:coverage} shows that the coverage in this case reduces by changing the benchmark from 1 to 2, and this reduction is significantly greater when log priors are imposed. Comparison of $\mhalf$ values with neutralino masses $m_{\tilde\chi^0_1}$ (Figs.~\ref{fig:2D_phys_priors_CMSSMmChi}c,d) indicates a correlation between the two, as expected in the CMSSM. This is not surprising though when we recall the actual correlation between $m_{\tilde\chi^0_1}$ and $\mhalf$ in the CMSSM. The thin straight line depicts the strong correlation between the two quantities, and this means that we expect to see very similar behaviours for the two cases in any statistical inference; these include variation patterns for the coverage. As an alternative way of explaining our observations for $\mhalf$, we could instead look directly at Figs.~\ref{fig:2D_phys_priors_CMSSMsigmaSI}c,d for $\mhalf$ versus $\sigma^{SI}_p$. Since the plots are very similar to the ones in~\fig{fig:2D_phys_priors_DD} for $m_{\tilde\chi^0_1}$ and $\sigma^{SI}_p$, our previous arguments hold here as well.

In order to understand our results in~\tab{tab:coverage} for the scalar mass parameter $\mzero$, consider Figs.~\ref{fig:2D_phys_priors_CMSSMsigmaSI}a,b for $\mzero$ versus $\sigma^{SI}_p$ in both flat and log priors. Let us first look at~\fig{fig:2D_phys_priors_CMSSMsigmaSI}b for log priors. The bias in the distribution of the sample points in the figure toward lower $\mzero$ and $\sigma^{SI}_p$ shows that the scanning algorithm tends to sample regions with lower values of these quantities. There are more sample points around benchmark 1 in this plane, meaning that the likelihood is scanned more accurately in the vicinity of benchmark 1 compared to benchmark 2. This observation explains the large decrease in the coverage for the latter case.~\fig{fig:2D_phys_priors_CMSSMsigmaSI}a for flat priors, shows similar patterns although they are not as strong as in~\fig{fig:2D_phys_priors_CMSSMsigmaSI}b. This is completely consistent with our numbers for $\mzero$ in~\tab{tab:coverage} when flat priors are imposed: coverage does drop from benchmark 1 to 2, but less substantially than in the log prior case.

Finally, for the remaining two CMSSM parameters, $\azero$ and $\tanb$,~\tab{tab:coverage} does not show large variations in coverage in different cases. These can again be explained from what we observe in Figs.~\ref{fig:2D_phys_priors_CMSSMmChi}e,f and Figs.~\ref{fig:2D_phys_priors_CMSSMmChi}g,h for $\azero$ and $\tanb$ versus $m_{\tilde\chi^0_1}$ in both flat and log priors. Plots show that our benchmarks 1 and 2 are not considerably far from each other in these planes, and are hence sampled approximately equally well. Effects on the likelihood reconstruction of moving from one point to the other are therefore not large, and result in only slight changes to the coverage. We also do not see any major correlations between these parameters and $\sigma^{SI}_p$ in Figs.~\ref{fig:2D_phys_priors_CMSSMsigmaSI}e,f and Figs.~\ref{fig:2D_phys_priors_CMSSMsigmaSI}g,h, confirming this interpretation.

As we stated several times in the above discussions, our interpretation of the poor coverage in our results is that the issue mainly stems from the sampling effects that are induced by the imperfection of the employed scanning algorithm in accurately reconstructing the profile likelihoods rather than the breakdown of the profile likelihood approximation of the frequentist confidence intervals. This means that by using an algorithm that is optimised for frequentist analysis of the parameter spaces, it should be possible to substantially alleviate the coverage issue in these cases.

\begin{table}[t]
\begin{center}
\begin{tabular}{|l | >{\centering\arraybackslash}p{1cm} | >{\centering\arraybackslash}p{2cm} | >{\centering\arraybackslash}p{2cm}|} \cline{3-4}
\multicolumn{2}{c|}{}& \multicolumn{2}{c|}{\textbf{Benchmark 2 (conf. int.)}} \bigstrut[t]\\ \cline{3-4}
\multicolumn{2}{c|}{} & $1\sigma$ & $2\sigma$ \bigstrut\\ \hhline{--==}
\multirow{8}{*}{\begin{sideways}\textbf{Flat + log priors}\end{sideways}} & $\mzero$           &  \textcolor{red}{\textbf{\textbf{35}}} & \textcolor{red}{\textbf{\textbf{72}}} \bigstrut \\ \cline{2-4}\cline{2-4}
& $\mhalf$           &  \textcolor{red}{\textbf{15}} & \textcolor{red}{\textbf{55}} \bigstrut\\ \cline{2-4}
& $\azero$           &  \textcolor{black}{\textbf{92}} & \textcolor{black}{\textbf{100}} \bigstrut\\ \cline{2-4}
& $\tanb$           &  \textcolor{black}{\textbf{100}} & \textcolor{black}{\textbf{100}} \bigstrut\\ \cline{2-4}
& $m_{\tilde\chi^0_1}$           &  \textcolor{red}{\textbf{23}} & \textcolor{red}{\textbf{60}} \bigstrut\\\cline{2-4}
& $\sigma^{SI}_p$           &  \textcolor{red}{\textbf{21}} & \textcolor{red}{\textbf{63}} \bigstrut\\ \hline
\end{tabular}
\caption[aa]{\footnotesize{As in~\tab{tab:coverage}, but when the sample points of the scans with flat and log priors are combined and the confidence intervals are constructed based on the total points. The table shows the combined analysis for benchmark 2 and confidence intervals only. Again, numbers in red and black show under- and over-coverage, respectively.}} \label{tab:flatpluslog}
\end{center}
\end{table}

If one wants to adhere to the Bayesian techniques such as the one we have employed here, one potential solution would be to first explore the parameter space using different types of priors and then construct the confidence intervals from the combination of all sample points that are obtained this way. This is because each set of priors leads the algorithm to scan different parts of the parameter space with different resolutions and consequently the combined sample points would provide a better mapping of the likelihood over the entire parameter space. We investigate this for the particular cases of the flat and log priors utilised in our scans. The procedure is the following: For every flat-prior scan we have performed, we randomly choose one of our 100 log-prior scans and then merge the two sample sets into a new one. We repeat this for all 100 flat-prior scans and in the end we obtain 100 new sets of sample points. We then construct the one-dimensional profile likelihoods for each set and study the degrees of coverage in exactly the same way as before. Our coverage results for benchmark 2 and confidence intervals are given in~\tab{tab:flatpluslog}. The numbers clearly show that combining the two sets of sample points does not improve the coverage in this case (compare with~\tab{tab:coverage}). For parameters $\mzero$, $\mhalf$, $m_{\tilde\chi^0_1}$ and $\sigma^{SI}_p$, where both flat- and log-prior cases show under-coverage in~\tab{tab:coverage}, we again observe severe under-coverage in the combined case. While the coverage improves compared to the case with log priors only, it deteriorates compared to the flat-prior case. We discuss a possible reason for this somewhat counter-intuitive result at the end of this section. For $\azero$ and $\tanb$ on the other hand, where both flat and log priors of~\tab{tab:coverage} show similar over-coverage, the combined results also show over-coverage with numbers very similar to the individual cases. Although our results indicate that here flat-prior scans provide better coverage than the log and (flat+log) scans, they suggest that for cases where flat and log priors give over- and under-coverage, respectively, their combination could result in a proper coverage.

Another way to improve the coverage when Bayesian scanning algorithms are used is to modify the configuration of the algorithms (if possible) so as to make them more optimised for frequentist analyses. In the coverage evaluations of this paper, we use~\MN~with a configuration that is widely used in SUSY parameter estimation (see e.g. Ref.~\cite{Trotta:08093792}): the number of live points, the tolerance parameter and the maximum efficiency are set to 4000, 0.5 and 1, respectively. This corresponds to around 50000 points in the final set of samples for each scan. After the initial submission of this paper however, a new configuration of~\MN~was suggested~\cite{Feroz:2011bj} that maps profile likelihoods from present-day data much more accurately. Here a smaller tolerance parameter (0.0004) and a significantly higher number of live points (20000) are used, and the algorithm requires substantially longer run-time. We have investigated in a few individual parameter scans whether running~\MN~with this new configuration leads to substantially different intervals for the profile likelihoods than the ones we have obtained in the present work. The results are however not conclusive. In some cases, we obtain larger contours for the confidence regions, suggesting that running with the new configuration might increase the coverage.  It is however possible that the changes we observe are solely due to sampling noise, rather than any real improvement in the coverage.

Although improving the mapping of the profile likelihood should typically result in improved coverage in situations where under- or over-coverage can be attributed to deficiencies in scanning techniques, this will also not necessarily be borne out in the results of every individual parameter scan.  This is because there are two competing effects in mapping the likelihoods with~\MN: (1) The scan with the new configuration might produce larger contours because it finds more points which lie above the threshold likelihood. (2) It might produce smaller contours, because it should find a higher best-fit point, shifting the thresholds for $1$- and $2\sigma$ up. This may also explain the curious result discussed above that merged (flat+log) prior scans sometimes show poorer coverage than simple flat prior scans. In order to conclusively determine whether changing the~\MN~parameters improves the coverage or not, one needs to perform a complete coverage study using the new configuration, which is well beyond the scope of the current paper given the massive computational resources required for such an investigation.

\section{Conclusions} \label{sec:concl}

We have investigated the ability of advanced Bayesian scanning techniques to correctly construct frequentist confidence intervals for supersymmetric parameter estimation. Our construction has been based on one-dimensional profile likelihoods obtained by maximising the full multi-dimensional likelihood function over all unwanted parameters. We chose the CMSSM for our analysis, giving us four free continuous parameters.  We added to this parameter space four SM quantities as nuisance parameters. We assessed correctness of the intervals by evaluating the statistical coverage for two benchmark points in the parameter space when either flat or logarithmic priors are imposed on the GUT-scale unified mass parameters $\mzero$ and $\mhalf$. We used a likelihood for direct detection of dark matter particles with a XENON10-like experiment. This gave us a rather high-statistics likelihood. We explored the parameter space with~\MN, a state-of-the-art scanning technique based on nested sampling and we set the configuration parameters of the algorithm to the commonly used values optimised for Bayesian inference.

Our results show poor coverage for most cases. We see both over- and under-coverage that vary for different benchmarks and priors. We studied effects of sampling biases caused by the explicit priors as well as the physicality constraints imposed on different regions of the parameter space, and showed that these effects can be very strong in some cases. This causes the sampling algorithm to explore different regions with different resolutions, which consequently impacts its ability to accurately map the likelihood function. Some regions with direct impact on the statistical inference and construction of the confidence intervals can be entirely missed. Our interpretations of the observed poor coverage (especially in under-coverage cases) were mainly based on these sampling effects, i.e. the inadequacy of the Bayesianally-optimised scanning algorithms in constructing profile likelihoods. Although poor coverage can arise alternatively from the breakdown of profile likelihood approximation of the full Neyman construction of confidence intervals, we did not consider this as the dominate effect in our analysis. This effect usually happens when the true parameters lie at the boundary of the accessible parameter space which for SUSY models is set by physicality constraints. Our benchmarks however did not seem to be located at the boundary. In addition, the boundary effects always lead to over-coverage, contrary to the fact that our results show also substantial under-coverage in various cases.

The sampling effects are expected to significantly weaken in situations where very high-statistics data are available. This could happen with, for example, collider data from the LHC or DM DD data from the future ton-scale experiments.

We have seen that imposing log priors does not help improve coverage, and that results even deteriorate in most cases. In addition, the confidence intervals constructed from the combination of flat- and log-prior samples do not show proper coverage. This means that if one wants to use the scanning techniques that are optimised for Bayesian inference to correctly estimate parameters of a complex model such as the CMSSM in a frequentist framework, other more sophisticated types of priors should be employed. Some Bayesian techniques, including~\MN, can be configured in such a way that become more appropriate for profile likelihood analyses. Utilising these different configurations should provide a substantially better coverage given that they map the likelihood accurately enough and the poor coverage is predominantly due to the sampling effects. It is however entirely possible for a complex parameter space that the profile likelihood does not approximate the full Neyman construction accurately enough, and consequently, the confidence intervals constructed based on it do not cover the true parameters properly. In these cases, one can replace the profile likelihood method with other more sophisticated construction methods, such as the one suggested by Feldman \& Cousins~\cite{Feldman:1998}.

Given that both MCMCs and nested sampling are optimised for calculating the Bayesian posterior PDF and not the profile likelihood, there is reason to suspect that the MCMC techniques employed in other frequentist scans of the CMSSM (e.g. Refs.~\cite{Buchmueller:2009fn,Buchmueller:2010ai,Buchmueller:2011aa}) might also show similarly poor coverage.  An investigation of the coverage properties of these techniques has yet to be presented, and would be a very interesting addition to the literature.

Our analysis has been limited to a very specific example of SUSY parameter estimation. The model is reduced to the CMSSM, the likelihood is restricted to a particular DD experiment and uncertainties in the halo model and neutralino-quark couplings are ignored (see e.g. Refs.~\cite{Akrami:2010,Strigari:2009zb} for the impacts of these uncertainties on the CMSSM parameter estimation). We made these choices particularly due to our limited computational power for generating the required pseudo-data. Ideally one should take into account likelihoods from all available astrophysical and collider experiments in a global-fit framework, including as many nuisance parameters as possible. Studying the coverage in these cases is of great importance for any frequentist inference on the model. The number of pseudo-experiments we used here was sufficient to demonstrate that there are issues with coverage for our particular chosen case.  This number could be increased in the future to obtain more accurate estimates of the coverage of such global fits. This however requires substantial computing resources and more efficient scanning techniques (for a coverage study of the CMSSM with ATLAS likelihoods and accelerated scans based on neural networks, see Ref.~\cite{Bridges2010}); we leave the investigation of such cases for future work. 

\acknowledgments{We thank the authors of Ref.~\cite{Bridges2010} for sharing their work with us before publication and for useful comments on this work.  We also thank an anonymous referee for helpful comments. We are grateful to the Swedish Research Council (VR) for financial support.  PS is supported by the Lorne Trottier Chair in Astrophysics and an Institute for Particle Physics Theory Fellowship.  JC is a Royal Swedish Academy of Sciences Research Fellow supported by a grant from the Knut and Alice Wallenberg Foundation. JE thanks the Swedish Research Council for support (contract 621-2010-3301).}


\end{document}